%% file: main.tex
\documentclass[twocolumn,svgnames]{svjour3}
\usepackage[utf8]{inputenc}
\usepackage[dvipsnames]{xcolor}
\usepackage{amsmath}
\usepackage{graphicx}
\usepackage{hyperref}
\usepackage{caption}

\usepackage{tikz}
\usetikzlibrary{chains,shadows.blur,fit,patterns}
\newcommand{\ignore}[1]{}
\usepackage{listings}
\usepackage{stmaryrd}
\usepackage{balance}

\usepackage{todonotes}

\usepackage{bm}

\makeatletter
\let\cl@chapter\undefined
\makeatletter

\usepackage[capitalise]{cleveref}
\usepackage{subcaption}
\usepackage{csvsimple}
\input{defs.tex}

\usepackage{enumitem}

\definecolor{lgreen} {RGB}{180,210,100}
\definecolor{norange}{RGB}{230,120,20}

\usepackage{letltxmacro}
\newcommand*{\SavedLstInline}{}
\LetLtxMacro\SavedLstInline\lstinline
\DeclareRobustCommand*{\lstinline}{%
  \ifmmode
    \let\SavedBGroup\bgroup
    \def\bgroup{%
      \let\bgroup\SavedBGroup
      \hbox\bgroup
    }%
  \fi
  \SavedLstInline
}

\journalname{{}}

\begin{document}
\date{\today}

\title{Interactive Abstract Interpretation:\\ Reanalyzing Whole Programs for Cheap}

\author{Julian Erhard         \and
        Simmo Saan \and
        Sarah Tilscher \and
        Michael Schwarz \and
        Karoliine Holter \and
        Vesal Vojdani \and
        Helmut Seidl
}

\authorrunning{J. Erhard et al.\ } 

\institute{J. Erhard, S. Tilscher, M. Schwarz, H. Seidl \at
              Technical University of Munich \\
              \email{ \{julian.erhard, sarah.tilscher, m.schwarz, helmut.seidl\}@tum.de}           
           \and
           S. Saan, K. Holter, V. Vojdani \at
              University of Tartu \\
              \email{ \{simmo.saan, karoliine.holter, vesal.vojdani\}@ut.ee}
}

\maketitle

\begin{abstract}
    To put static program analysis at the fingertips of the software developer,
    we propose a framework for interactive abstract interpretation.
    While providing sound analysis results, abstract interpretation in general can be quite costly.
    To achieve quick response times, we incrementalize the analysis infrastructure, including postprocessing,
    without necessitating any modifications to the analysis specifications themselves.
    We rely on the local generic fixpoint engine \textbf{TD}, which dynamically tracks dependencies,
    while exploring the unknowns contributing to answering an initial query.
    Lazy invalidation is employed for analysis results affected by program change.
    Dedicated improvements support the incremental analysis of
    concurrency deficiencies such as data-races.   %
    The framework has been implemented for multithreaded C within the
    static analyzer \textsc{Goblint}, using \textsc{MagpieBridge} to relay findings to IDEs.
    We evaluate our implementation w.r.t.\ the yard sticks of response time and consistency: formerly proven invariants
    should be retained -- when they are not affected by the change.
    The results indicate that with our approach, a reanalysis after small changes only takes a fraction of from-scratch analysis time,
    while most of the precision is retained.
    We also provide examples of program development highlighting the usability of the overall approach.
\end{abstract}

\input{content/01-intro.tex}
\input{content/02-background.tex}
\input{content/03-reluctant.tex}
\input{content/04-restart.tex}
\input{content/05-pipeline.tex}
\input{content/06-practical.tex}
\input{content/07-consistency.tex}
\input{content/08-evaluation.tex}
\input{content/09-related.tex}
\input{content/10-conclusion.tex}

\paragraph{Acknowledgements.} This work was supported by the German Research Foundation (DFG) - 378803395/2428 \textsc{ConVeY} and the Estonian Research Council - PSG61.

\bibliographystyle{spmpsci}
\balance
\bibliography{lit}



\end{document}

%% file: content/01-intro.tex
\section{Introduction}
\label{sec:introduction}

\begin{figure*}[t]
    \resizebox{\textwidth}{!}{
        \hspace*{-2.6cm} 
        \input{content/fig-pipeline.tex}
    }
    \caption{An overview of the interactive analysis pipeline, showing how previous analysis results are used in the analysis of modified code to yield new analysis results. The shaded components and trivial postprocessing make up the corresponding non-incremental analyzer.}
    \label{f:pipeline}
\end{figure*}
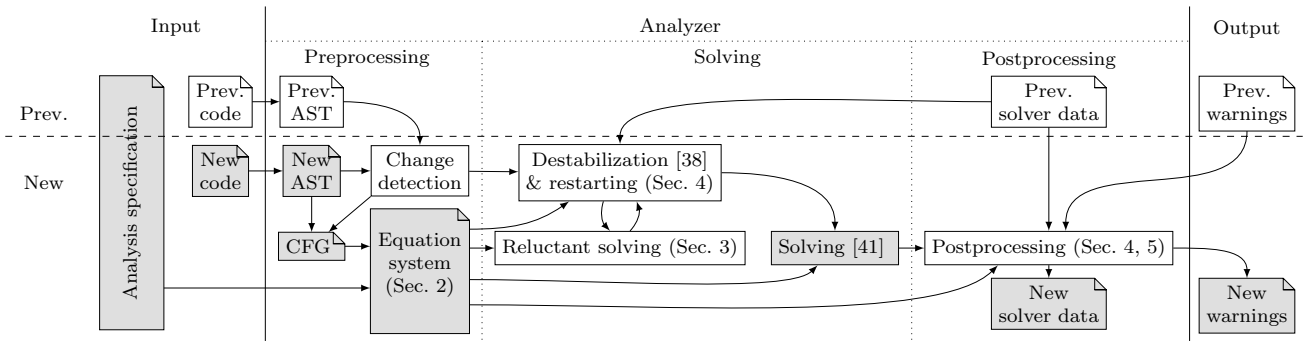

Heeding the call to ``dramatically reduce software vulnerability''~\cite{NIST2016},
a variety of static analysis tools have been developed, e.g.\ \cite{coverity,findbugs,cppcheck,infer,astree,Goblint2016,framac,CPAchecker}.
Several tools provide users with hints on potential bugs \cite{coverity,findbugs,cppcheck} -- without giving any guarantees that absence of warnings
implies absence of programming errors.
For critical applications, however, \emph{sound} analyses are important to prove the absence of particular classes of errors. Classical examples are
absence of null pointer dereferences, buffer overruns or data-races.
One general framework for constructing sound static analyses is
\emph{abstract interpretation}~\cite{DBLP:conf/popl/CousotC77}.
Several static analysis tools, for instance \cite{framac,astree,Goblint2016,mine2018mopsa}, build on this framework.
The usage of sound static analysis tools by developers, however, is not as ubiquitous as one might hope for.
To investigate the reasons for the lacking adoption of (sound and unsound) static analysis tools, qualitative developer surveys such as \cite{DBLP:conf/icse/JohnsonSMB13,DBLP:conf/kbse/ChristakisB16} were conducted.
Slow runtime and lack of integration into the development process were among the most common points of criticism mentioned by developers that had experience
in using static analysis software.
Other criticism included poor phrasing of warnings generated by these tools and a high false-positive rate, i.e., a too large proportion of warnings not reflecting actual programming errors.
%

%
%
Here, we address the first two issues. We present enhancements to abstract interpretation based analyses to make them suited for interactive usage,
i.e., allow for a seamless integration into the software development process.
%
%
We consider the issues of reducing the number of false-positives, as well as of properly phrasing warnings
as equally important, yet orthogonal questions.

To tackle the problem of lacking performance, \emph{incremental} analyses that reuse analysis results where possible have been proposed.
One approach to incremental analysis relies on summaries to modularize the analysis effort~\cite{Relay,DBLP:conf/lics/OHearn18,DBLP:conf/sigsoft/McPeakGR13,Calcagno11Abduction}.
%
\emph{Sound} modular analyses, however, suffer from the necessity to summarize the effects for every conceivable calling or execution context.
Accordingly, these summaries must necessarily either be rather coarse or very expensive.
An interesting approach is
presented in \cite{CousotCFL13} where soundness is traded for an improved false-positivity rate.
There, the goal is to find \emph{necessary} pre-conditions, i.e., pre-conditions that, when violated,
are guaranteed to lead to errors. By construction, such an approach cannot be sound as well.


%
On the other hand, the approach of \cite{Seidl2020} considers a \emph{whole-program} analyzer based on a particular generic fixpoint engine
solver \cite{DBLP:conf/ppdp/SeidlV18,seidl_vogler_2022}, and
points out that the data-structures already provided by the solver can be exploited
to \emph{limit} the amount of reanalysis after a modification of the code. In this way, not only the modified code snippet,
but also those parts of the program which are affected by the change, are reanalyzed.
The resulting incremental analyzer can be fully or partially context-sensitive and takes
concurrency of the program into account \cite{DBLP:conf/aplas/ApinisSV12} -- which may be hard in a purely modular approach.
This approach should be contrasted to the incremental
whole-program analysis of Stein et al. \cite{Stein21}, Stein \cite{stein_phd}.
There, acyclic \emph{demanded abstract interpretation graphs} are used to encode the abstract interpretation process.
This approach requires dedicated constructions for dealing with loops, dynamic function calls or recursion.
To the best of our knowledge, multithreaded code is still out of reach.



In contrast, the goal of the present paper is to develop
an \emph{interactive} analyzer
for multi\-threaded C programs which does not require any dedicated extensions to deal with specific program structures.
%
This analyzer should be integrated into commonly used IDEs. After every modification of the code, it should
provide the developer with an updated set of warnings displayed in the IDE itself reducing the number of mental context switches
the developer has to make.

We identify two yardsticks for evaluating the practicality of interactive static analysis, namely,
\emph{efficiency} and \emph{consistency}. The first criterion requires that reanalysis after minor modifications
of the program should be considerably faster than a
from scratch non-incremental analysis of the code base,
while the second asks for an incremental run to provide analysis results that match those obtained by a from-scratch reanalysis as closely as possible.
We remark that reanalyzing only parts of the code may indeed come at the price
of deteriorating the precision of results, namely, when parts of the old analysis result are extensively reused.
In particular, this loss in precision concerns data collected in a flow-insensitive manner.
%
Similar yardsticks have been defined in \cite{DBLP:conf/sigsoft/McPeakGR13}, namely being \emph{fast} and \emph{deterministic}.
In their setting of parallelized (possibly unsound) bug hunting, determinism of the results cannot be taken for granted.
Since we do not (yet) consider parallelization of the analyzer, determinism is irrelevant, while
consistency is of major importance as it measures the deterioration of analysis results
across multiple reanalyses.
%
In order to make the abilities of an incremental analyzer available to the developer,
we build on \textsc{MagpieBridge} \cite{luo_et_al:LIPIcs:2019:10813} to integrate the
feedback provided by the analyzer into the developer's favorite IDE.
The communication between \textsc{MagpieBridge} and the respective IDEs is then done via the Language Server Protocol~\cite{LSP} and HTTP.
Functioning as an adapter, \textsc{MagpieBridge} offers a single interface to target, while aiming at a wide range of IDE support.
In detail, our contributions are:

\begin{itemize}
\item Several improvements to the incremental analyzer from \cite{Seidl2020}, namely,
	\begin{itemize}
	\item	\emph{reluctant} destabilization to possibly confine the reanalysis to the bodies of modified functions;
	\item	a \emph{restarting} mechanism for precision improvement when flow-insensitive invariants about\linebreak[4] global data-structures are affected;
	\item   a \emph{cheap} restarting for a benign class of flow-insensitive information;
	\end{itemize}
\item Incrementalization not only of the analysis itself, but also of the necessary postprocessing;
\item A reanalysis infrastructure which may interface  with various IDEs by means of \textsc{MagpieBridge} \cite{luo_et_al:LIPIcs:2019:10813};
\item A systematic evaluation of the given improvements on a data-race analysis of
      not-too-small real-world benchmarks w.r.t.\ the yardsticks \emph{efficiency} and \emph{consistency}.
\end{itemize}
\noindent
The rest of the paper is structured as follows: \Cref{sec:background} gives a summary of the local solver and the work we build upon.
\cref{sec:reluctant} is dedicated to how the incremental analysis can be made even more efficient, using \textit{reluctant destabilization},
which employs deferring the destabilization of unknowns influenced by changed functions until it is clear that it is necessary.
\cref{sec:restart} describes how the precision of the incremental analysis can be improved via restarting.
In \cref{sec:implementation} we give insights on the key challenges faced when implementing the proposed interactive analysis in a static analysis tool.
The experimental evaluation can be found in \cref{sec:evaluation}.
\cref{sec:related} gives a brief overview on more recent related work on incremental analysis as well as related work on the restarting of unknowns to improve analysis precision.
\cref{sec:conclusion} concludes and gives an outlook on future work.

%% file: content/fig-pipeline.tex
\usetikzlibrary{shapes.geometric,matrix,fit,calc}

\makeatletter
\pgfdeclareshape{document}{
    \inheritsavedanchors[from=rectangle] 
    \inheritanchorborder[from=rectangle]
    \inheritanchor[from=rectangle]{center}
    \inheritanchor[from=rectangle]{north}
    \inheritanchor[from=rectangle]{south}
    \inheritanchor[from=rectangle]{west}
    \inheritanchor[from=rectangle]{east}
    \inheritanchor[from=rectangle]{base}
    \backgroundpath{
        \southwest \pgf@xa=\pgf@x \pgf@ya=\pgf@y
        \northeast \pgf@xb=\pgf@x \pgf@yb=\pgf@y
        \pgf@xc=\pgf@xb \advance\pgf@xc by-5pt 
        \pgf@yc=\pgf@yb \advance\pgf@yc by-5pt
        \pgfpathmoveto{\pgfpoint{\pgf@xa}{\pgf@ya}}
        \pgfpathlineto{\pgfpoint{\pgf@xa}{\pgf@yb}}
        \pgfpathlineto{\pgfpoint{\pgf@xc}{\pgf@yb}}
        \pgfpathlineto{\pgfpoint{\pgf@xb}{\pgf@yc}}
        \pgfpathlineto{\pgfpoint{\pgf@xb}{\pgf@ya}}
        \pgfpathclose
        \pgfpathmoveto{\pgfpoint{\pgf@xc}{\pgf@yb}}
        \pgfpathlineto{\pgfpoint{\pgf@xc}{\pgf@yc}}
        \pgfpathlineto{\pgfpoint{\pgf@xb}{\pgf@yc}}
        \pgfpathlineto{\pgfpoint{\pgf@xc}{\pgf@yc}}
    }
}
\makeatother

    \begin{tikzpicture}[
            data/.style={
                draw,document,align=center
            },
            proc/.style={
                draw,rectangle,align=center
            },
            font=\small,
            >=latex,
            every fit/.style={
                inner sep=0,
            },
            nonincr/.style={
                fill=lightgray,
                fill opacity=0.5,
                text opacity=1
            },
            contrib/.style={
            }
        ]
        \matrix (mat) [matrix of nodes,nodes in empty cells=false,column sep=3ex]
        {
            & |(inputleft)| & |(inputright)| & |(analyzerleft)| & & & & |(analyzerright)| & |(output)| Output \\
            & & & |(preleft)| & |(preright)| & |(solleft)| & |(solright)| & Postprocessing & \\
            Prev. & \node[align=center] (specup) {\phantom{Prev.} \\ \phantom{}}; & \node[data] (prevcode) {Prev. \\ code}; & \node[data] (prevast) {Prev. \\ AST}; & & & & \node[data] (prevsol) {Prev. \\ solver data}; & \node[data] (prevwarn) {Prev. \\ warnings}; \\[1.5ex]
            New & & \node[data,nonincr] (newcode) {New \\ code}; & \node[data,nonincr] (newast) {New \\ AST}; & \node[proc] (changed) {Change \\ detection}; & \node[proc,contrib] (destab) {Destabilization~\cite{Seidl2020} \\ \& restarting~(Sec.~\ref{sec:restart})}; \\[1ex]
            & & & |[data,nonincr] (cfg)| CFG & \node[align=center] (eqsysup) {\phantom{Equation} \\ \phantom{system}}; & \node[proc,contrib] (relsolve) {Reluctant solving~(Sec.~\ref{sec:reluctant})}; & |[proc,nonincr] (solve)| Solving~\cite{seidl_vogler_2022} & \node[proc,contrib] (wo) {Postprocessing~(Sec.~\ref{sec:restart}, \ref{sec:pipeline})}; \\[1.5ex]
            & \node[align=center] (specdown) {\phantom{New} \\ \phantom{}}; & & & \node[align=center] (eqsysdown) {\phantom{Equation} \\ \phantom{system}}; & & & \node[data,nonincr] (newsol) {New \\ solver data}; & \node[data,nonincr] (newwarn) {New \\ warnings}; \\
        };
        \node[fit=(inputleft) (inputright)] (input) {Input};
        \node[fit=(preleft) (preright)] (pre) {Preprocessing};
        \node[fit=(solleft) (solright)] (sol) {Solving};
        \node[fit=(analyzerleft) (analyzerright)] (analyzer) {Analyzer};
        \node[data,nonincr,fit=(eqsysup) (eqsysdown)] (eqsys) {Equation \\ system \\ (Sec.~\ref{sec:background})};
        \node[data,nonincr,fit=(specup) (specdown)] (spec) {\rotatebox[origin=rB]{90}{Analysis specification}};
        \path[->] ([yshift=-2ex] spec.east |- eqsys) edge ([yshift=-2ex] eqsys.west);
        \path[->] (prevcode) edge (prevast);
        \path[->] (newcode) edge (newast);
        \path[->] (newast)
            edge (cfg)
            edge (changed);
        \path[->] (prevast) edge [out=0,in=90] (changed);
        \path[->] (cfg) edge (eqsys.west |- cfg);
        \path[->] (changed) edge (destab)
            (changed.south west) edge (cfg);
        \path[->]
            ([yshift=5ex] eqsys.east) edge [out=0,in=-145] ([xshift=-6ex] destab.south)
            (eqsys.east |- relsolve) edge (relsolve)
            ([yshift=-1ex] eqsys.east) edge [out=0,in=-135,in looseness=0.3] (solve)
            ([yshift=-4ex] eqsys.east) edge [out=0,in=-135,in looseness=0.4] ([xshift=-6ex] wo.south);
        \path[->] (destab)
            edge [bend right] (relsolve)
            edge [out=0,in=90] (solve);
        \path[->] (relsolve) edge [bend right] (destab);
        \path[->] (solve) edge (wo);
        \path[->] (wo)
            edge (newsol)
            edge [out=0,in=90] (newwarn);
        \path[->] (prevsol)
            edge [out=180,in=90,in looseness=0.4] (destab)
            edge (wo);
        \path[->] (prevwarn) edge [out=-90,in=90] ([xshift=2ex] wo.north);
        
        \coordinate (input-pre) at ($(prevcode.east)!0.5!(cfg.west)$);
        \coordinate (pre-sol) at ($(eqsys.east)!0.5!(relsolve.west)$);
        \coordinate (sol-post) at ($(solve.east)!0.5!(wo.west)$);
        \coordinate (post-output) at ($(wo.east)!0.5!(output.west)$);
        \coordinate (analyzer-sol) at ($(analyzer.south)!0.5!(sol.north)+(0,1ex)$);
        \path[draw]
            (input-pre |- mat.north) to (input-pre |- mat.south)
            (post-output |- mat.north) to (post-output |- mat.south);
        \path[draw,dotted]
            (pre-sol |- analyzer-sol) to (pre-sol |- mat.south)
            (sol-post |- analyzer-sol) to (sol-post |- mat.south)
            (input-pre |- analyzer-sol) to (post-output |- analyzer-sol);
        
        \coordinate (prev-new) at ($(prevcode.south)!0.5!(newcode.north)$);
        \path[draw,dashed] (mat.west |- prev-new) to (mat.east |- prev-new);
    \end{tikzpicture}

%% file: content/02-background.tex
\section{Preliminaries \label{sec:background}}

In this section, we sketch the basic structure of the non-incremental static analyzer we base our work on (see \cref{f:pipeline}).
Conceptually, this analyzer represents the program to be analyzed as a system of equations
$x = f_x$, where the unknowns $x\in\mathcal{X}$, possibly occurring on left-hand sides,
correspond to program points (possibly decorated with calling-contexts) or global variables
for which invariants are to be determined.
The invariants to be computed are represented by means of \emph{abstract} values from
some \emph{complete lattice}
(opposed to the \emph{concrete} program states as observed during an
actual run of the program).

This base setting is extended such that during the evaluation of a right-hand side $f_x$, contributions
to further unknowns $g$ may also be produced as \emph{side-effects}, resulting in
\emph{side-effecting} systems of equations \cite{seidl2003global,DBLP:conf/aplas/ApinisSV12,seidl_vogler_2022}.
This mechanism is particularly useful to accumulate flow-insensitive information
(such as an invariant of  some data-structure shared among multiple threads),
during a flow- and context-sensitive analysis of procedural code.

\begin{example}\label{e:side}
Consider the example program from \cref{f:side} together with its control-flow graph, and assume that we perform
a value set analysis for each program variable. The set of unknowns then consists of a single unknown for the
global variable $g$ taking as values sets \textbf{int}s.
Additionally, the analysis maintains the set of all pairs
$\angl{3,\alpha},\angl{4,\alpha},\angl{5,\alpha}$ for calling contexts $\alpha$ of the \textsf{main} function, and
the set of all pairs
$\angl{0,\beta},\angl{1,\beta},\angl{2,\beta}$ for calling contexts $\beta$ of the function \textsf{foo}.
Each of these unknowns may take as values assignments of local variables to sets of values of corresponding types.
Since the parameter list of \textsf{main} is empty, $\alpha$ is given by the empty assignment $\emptyset$,
while the calling contexts of \textsf{foo} consist of all mappings $\{p\to S\}$ where $S$ is a set of addresses.

In general, the resulting system of equations thus will be huge, if not infinite.
We observe, however, that
when querying the analysis result for the \emph{endpoint} $5$ of \textsf{main} (in calling context $\emptyset$),
the abstract values of unknowns for the program points of \textsf{foo} are required for the single calling context
$\beta = \{p\mapsto\{\&g\}\}$.
For these, the right-hand side functions for the program points of \textsf{foo} are given by:
\[
\begin{array}{lll}
f_{\angl{0,\{p\mapsto\{\&g\}\}}}\,\rho	&=& \bot	\\
f_{\angl{1,\{p\mapsto\{\&g\}\}}}\,\rho	&=& \textbf{let}\;\sigma = \rho\,\angl{0,\{p\mapsto\&g\}}\,p) \\
					& & \textbf{in}\;\textbf{forall}\,(\&h\in\sigma\,p)\; \textsf{side}\,h\,\{1\};\\
					& & \phantom{\textbf{in}}\;\sigma	\\
f_{\angl{2,\{p\mapsto\{\&g\}\}}}\,\rho	&=& \rho\,\angl{1,\{p\mapsto\{\&g\}\}}\oplus\{\textsf{ret}\mapsto\{\textsf{null}\}\}
\end{array}
\]
Here, the operator ``$\oplus$'' updates the variable assignment in the left argument with the bindings provided on the right.
The right-hand side function for program point~1 produces the contribution $\{1\}$ as a side-effect
to all unknowns corresponding to globals that $p$ may point to (in this case, just $g$) by calling the external function \textsf{side},
before returning the local program state of the predecessor program point $0$
within the same calling context $\{p\mapsto\{\&g\}\}$.
The right-hand side function for program point 2 updates the program state of its predecessor
program point~1 by setting the value of \textsf{ret} to the set $\{\textsf{null}\}$
(assuming that the local variable \textsf{ret} receives the value to be returned by a call).
%
The right-hand side function for the start point of \textsf{foo} on the other hand, provides
the abstract value $\bot$ (representing the \emph{empty set} of program states).
The idea is that a non-$\bot$ abstract value may arrive at the start point of the function \textsf{foo} only
via a side-effect from some call or thread creation site of \textsf{foo}.
Such a side-effect occurs within the function \textsf{main} at program point 4:
\[
{\small
\begin{array}{lll}
f_{\angl{4,\emptyset}}\,\rho	&=& \textbf{let}\;\_=\textsf{side}\,\angl{0,\{p\mapsto\{\&g\}\}}\,
					\{\textsf{ret}\mapsto\top,\textsf{p}\mapsto\{\&g\}\}	\\
				& & \textbf{in\;let}\;\_ = \rho\,\angl{2,\{p\mapsto\{\&g\}\}}	\\
				& & \textbf{in}\;\rho\,\angl{3,\emptyset}
\end{array}}
\]
Here, the abstract value $\top$ represents the set of all possible values of the given type.
Note that in order to trigger the analysis of the function \textsf{foo} executed by the created thread,
the abstract value attained at the
end point of \textsf{foo} is explicitly queried (and then ignored) within the right-hand side function
for thread creation.
The right-hand side functions for the remaining program points of \textsf{main} are
\[
\begin{array}{lll}
f_{\angl{3,\emptyset}}\,\rho	&=& \bot \\
f_{\angl{5,\emptyset}}\,\rho	&=& \rho\,\angl{4,\emptyset}\oplus\{\textsf{ret}\mapsto\rho\,g\}
\end{array}
\]
In order to conveniently initialize globals,
we introduce a
dedicated main function \textsf{\_\_main} defined by
\begin{lstlisting}[basicstyle=\footnotesize]
void __main() {
   int ret;
   init();
   ret = main();
   return ret;
}
\end{lstlisting}
where \textsf{init} is a procedure which performs the initialization of globals
as specified by the program. In our case, it sets the global $g$ to 0.
In contrast to the start points of \textsf{foo} and \textsf{main}, the start point
of \textsf{\_\_main} cannot receive a non-$\bot$ value via a side-effect. Therefore,
it is provided with
a dedicated right-hand side providing the initial abstract value.
In our example with the single local variable \textsf{ret}, this abstract value is given by $\{\textsf{ret}\mapsto\top\}$
(unknown value). \qed

%
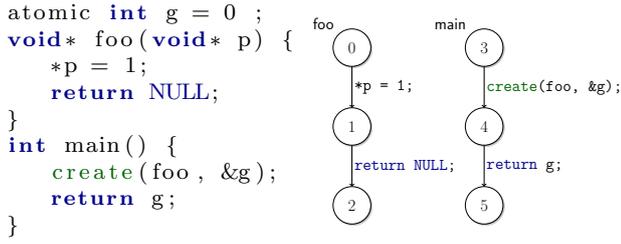
\begin{figure}
\hspace{-0.03\textwidth}
\begin{minipage}{0.22\textwidth}
\begin{lstlisting}[basicstyle=\footnotesize]
atomic int g = 0 ;
void* foo(void* p) {
   *p = 1;
   return NULL;
}
int main() {
   create(foo, &g);
   return g;
}
\end{lstlisting}
\end{minipage}%
\begin{minipage}{0.04\textwidth} 
\hspace{0.06\textwidth}
\end{minipage}
\begin{minipage}{0.24\textwidth}
\resizebox{1\textwidth}{!}{
   \begin{tikzpicture}[auto,
      node distance = 22mm,
      start chain = going below,
      f/.style = {font = \Huge},
      ed/.style = {f, ultra thick},
      state/.style = {draw,circle,blur shadow, fill=white,
            on chain,align=center, minimum size=2cm, f},
      stable/.style = {fill=lgreen}]

  \node[state] (foo1)    {$0$};
  \node[state] (foo2)    {$1$};
  \node[state] (foo3)    {$2$};
  \node (foo_label)[above left=0.2cm of foo1,f] {\textsf{foo}};

  \node[state] (main1)[right=5cm of foo1]    {$3$};
  \node[state] (main2)    {$4$};
  \node[state] (main3)    {$5$};
  \node (main_label)[above left=0.2cm of main1,f] {\textsf{main}};

  \begin{scope}[rounded corners]
   \path [->] (foo1) edge[ed] node {\lstinline{*p = 1;}} (foo2);
   \path [->] (foo2) edge[ed] node {\lstinline{return NULL;}} (foo3);

   \path [->] (main1) edge[ed] node {\lstinline{create(foo, \&g);}} (main2);
   \path [->] (main2) edge[ed] node {\lstinline{return g;}} (main3);
  \end{scope}
 \end{tikzpicture}
}
\end{minipage}
\caption{\label{f:side}A multithreaded program with its control-flow graph.}
\end{figure}
\end{example}
Generalizing the idea from \cref{e:side},
Schwarz et al.\ \cite{DBLP:conf/sas/SchwarzSSAEV21}
have used side-effecting equation systems to formalize several state-of-the-art
\emph{thread-modular} analyses of multithreaded code.
Thread-modular here means that an explicit consideration of all possible thread-\linebreak[4]interleavings is avoided.

Tackling the issue of large and possibly infinite equation systems, \emph{local} solvers, such as the top-down solver \cite{topdown1992},
have been proposed which do not return results for every unknown in the system, but only for those
\emph{contributing} to the result for an initially queried unknown.

\begin{example}\label{e:side2}
Consider, e.g., the program from \cref{f:side}.
Starting from a query to the unknown $\angl{5,\emptyset}$ corresponding to the end point
of the function \textsf{main} (or \textsf{\_\_main}) and the initial calling context, a local solver should query unknowns
corresponding to program points of \textsf{foo}, only for the calling context $\{p\mapsto\{\&g\}\}$.
%
%
As a solution $\rho$ for this system it may report
$\rho\,g = \{0,1\}$ together with
\[
\begin{array}[t]{lll}
\rho\,{\angl{0,\{p\mapsto\{\&g\}\}}}	&=& \{\textsf{ret}\mapsto\top,p\mapsto\{\&g\}\} \\
\rho\,{\angl{1,\{p\mapsto\{\&g\}\}}}	&=& \{\textsf{ret}\mapsto\top,p\mapsto\{\&g\}\} \\
\rho\,{\angl{2,\{p\mapsto\{\&g\}\}}}	&=& \{\textsf{ret}\mapsto\{\textsf{null}\},p\mapsto\{\&g\}\} \\[1ex]
%
\rho\,\angl{3,\emptyset}	&=& 	\{\textsf{ret}\mapsto\top\}	\\
\rho\,\angl{4,\emptyset}	&=& 	\{\textsf{ret}\mapsto\top\}	\\
\rho\,\angl{5,\emptyset}	&=& 	\{\textsf{ret}\mapsto\{0,1\}\}
\end{array}
\]
while all other unknowns have values $\bot$.
%
\qed
\end{example}
\noindent
We build on the \emph{top-down} solver $\mathbf{TD_{\emph{side}}}$ as presented in \cite{seidl_vogler_2022}.
This local solver extends the corresponding solver used in \cite{Seidl2020} by an explicit treatment of
side-effects occurring in right-hand sides as used, e.g., in our example.
An analysis of a program proceeds by calling $\mathbf{TD_{\emph{side}}}$ for a given unknown of interest which,
for C programs, is the endpoint of the function \textsf{main} in the initial calling context.

Besides a (partial) map from unknowns to their respective abstract values in some complete lattice $\D$,
the top-down solver maintains
the set \textsf{stable} of unknowns for which the fixpoint iteration has stabilized,
and a mapping \textsf{infl} to record for each encountered unknown $x$ the
set of other unknowns possibly influenced by the value of $x$.
%
%
In particular, it maintains the following data-structures:
\[
\begin{array}{lll}
\rho	&:&	(\X,\D)\,\textsf{map}	\\
\textsf{stable}	&:& \X\,\textsf{set}	\\
\textsf{infl}	&:& (\X,\X\,\textsf{set})\,\textsf{map}
\end{array}
\]
Beyond that, we let the solver record the mappings:
\[
\begin{array}{lll}
\textsf{side\_dep}	&:& (\X,\X\,\textsf{set})\,\textsf{map}	\\
\textsf{side\_infl}	&:& (\X,\X\,\textsf{set})\,\textsf{map}
\end{array}
\]
While the map \textsf{side\_dep} records for each global~$g$ the set of unknowns in whose right-hand side a side-effect to $g$ occurred,
the map \textsf{side\_infl} records for each unknown $x$ the set of globals to which the last evaluation of the right-hand side of $x$
triggered a side-effect.
The two maps are essentially inverses of each other.

\begin{example}\label{e:side3}
For the analysis from \cref{e:side} on the program from \cref{f:side}, the local solver
$\textbf{TD}_{\emph{side}}$ calculates the dependencies
\[
\begin{array}{|l|l|}
\hline
 &\textsf{infl}	\\
\hline
\textsf{init}			&	\{\_\_\textsf{main}\}		\\
g				&	\{\angl{5,\emptyset}\}	\\
\angl{0,\{p\mapsto\{\&g\}\}}	&	\{\angl{1,\{p\mapsto\{\&g\}\}}\}	\\
\angl{1,\{p\mapsto\{\&g\}\}}	&	\{\angl{2,\{p\mapsto\{\&g\}\}}\}	\\
\angl{2,\{p\mapsto\{\&g\}\}}	&	\{\angl{4,\emptyset}\}		\\
\angl{3,\emptyset}		&	\{\angl{4,\emptyset}\}		\\
\angl{4,\emptyset}		&	\{\angl{5,\emptyset}\}		\\
\angl{5,\emptyset}		&	\{\_\_\textsf{main}\}\\
\hline
\noalign{\vskip 6pt}
\hline
 &\textsf{side\_dep}	\\
\hline
g				&	\{\textsf{init}, \angl{1,\{p\mapsto\{\&g\}\}}\}	\\
\angl{0,\{p\mapsto\{\&g\}\}}	&	\{\angl{4,\emptyset}\}		\\
\angl{3,\emptyset}	&	\{\_\_\textsf{main}\}		\\
\hline
\noalign{\vskip 6pt}
\hline
 &\textsf{side\_infl}	\\
\hline
\textsf{init}		&	\{g\}		\\
\angl{1,\{p\mapsto\{\&g\}\}}	&	\{g\}	\\
\angl{4,\emptyset}		&	\{\angl{0,\{p\mapsto\{\&g\}\}}\}		\\
\_\_\textsf{main}		&	\{\angl{3,\emptyset}\}			\\
\hline
\end{array}
\]
where all empty set entries in tables have been omitted.~\qed
\end{example}

We remark that the implementation of the analyzer does not need to
construct an explicit representation of the complete constraint-system.
Rather, the transfer functions are constructed and evaluated on-demand using the control-flow graphs of the analyzed functions.

Subsequently, we present two enhancements of the local solver $\textbf{TD}_{\emph{side}}$
in order to arrive at an interactive analysis framework.

%% file: content/03-reluctant.tex
\section{Reluctant Destabilization} \label{sec:reluctant}

\paragraph{Base incremental setting}
When a program is modified, the updated system of equations differs from the old system in that a set $A$ of unknowns has received new right-hand side functions
(possibly introducing further new unknowns with corresponding right-hand sides).
The approach taken in \cite{Seidl2020} therefore is to remove the unknowns from set $A$ from the
set \textsf{stable} and then to eagerly call for $x\in A$ the function  \textsf{destabilize} from \cref{l:destabilize} in order to additionally
remove all unknowns from \textsf{stable} which directly or indirectly are influenced by unknown from $A$.
The granularity at which changes to the program are detected, is at the level of function definitions.
This means when a function is changed, all program points within it, except their start- and return-nodes are considered new, and so are the corresponding unknowns. The destabilization is then done at the return-nodes of these changed functions.
After this preprocessing, the solver is called for the initial unknown of interest $\angl{r_{main}, d_0}$, corresponding to the endpoint of the function \textsf{main}, in some particular calling context.
As the preprocessed data-structures are reused, this reanalysis will not iterate from scratch, but
take for each unknown $x$ the value into account which has
been computed for $x$ w.r.t.\ the old version of the equation system.
In particular, this means for those unknowns found in the set \textsf{stable},
that their values are looked up -- without any further re-evaluation.
For unknowns not yet in \textsf{stable}
the solver $\textbf{TD}_{\emph{side}}$ just continues with the iteration once they are encountered.

\begin{lstlisting}[language={[Objective]Caml}, label=l:destabilize, float, caption={Function destabilize.}]
destabilize x = let w = infl x in
		let infl x = |$\emptyset$| in
		forall (y|$\in$|w) {
		   stable -= y;
		   destabilize y;
		}
\end{lstlisting}

\begin{example}\label{e:side_non_reluctant}
Consider the program from \cref{f:side} from the last section,
and assume that in function \textsf{foo} the developer has changed the first assignment to
\begin{lstlisting}
*p = 2;
\end{lstlisting}
The approach from \cite{Seidl2020} will detect \textsf{foo} as changed,
create new identifiers for program nodes in this function, and call \textsf{destabilize} on all unknowns relating to its return-node.
The resulting stable set is visualized in \cref{fig:incr_destab}.
Besides the unknowns relating to nodes in the changed function \textsf{foo}, also the unknowns $\angl{4,\emptyset}$ and $\angl{5,\emptyset}$ are removed from \textsf{stable}.
Their values have to be re-computed once the solvers encounters them. \qed

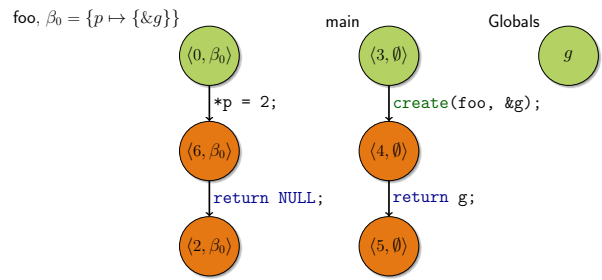
\begin{figure}
	\resizebox{0.45\textwidth}{!}{
		\begin{tikzpicture}[auto,
			node distance = 12mm,
			start chain = going below,
			f/.style = {font = \LARGE},
			ed/.style = {f, ultra thick},
			state/.style = {draw,circle,blur shadow,fill=norange,
					on chain,align=center, minimum size=2cm, f},
			stable/.style = {fill=lgreen}]

		\node[state, stable] (foo1)    {$\angl{0,\beta_0}$};
		\node[state] (foo2)    {$\angl{6,\beta_0}$};
		\node[state] (foo3)    {$\angl{2,\beta_0}$};
		\node (foo_label)[above left=0.2cm of foo1, f] {\textsf{foo}, $\beta_0=\{p\mapsto\{\&g\}\}$};

		\node[state,stable] (main1)[right=4cm of foo1]    {$\angl{3, \emptyset}$};
		\node[state] (main2)    {$\angl{4, \emptyset}$};
		\node[state] (main3)    {$\angl{5, \emptyset}$};
		\node (main_label)[above left=0.2cm of main1, f] {\textsf{main}};

		\node[state, stable] (g)[right=4cm of main1] {$g$};
		\node (global_label)[above left=0.2cm of g, f] {\textsf{Globals}};

		\begin{scope}[rounded corners]
		\path [->] (foo1) edge[ed] node {\lstinline{*p = 2;}} (foo2);
		\path [->] (foo2) edge[ed] node {\lstinline{return NULL;}} (foo3);

		\path [->] (main1) edge[ed] node {\lstinline{create(foo, \&g);}} (main2);
		\path [->] (main2) edge[ed] node {\lstinline{return g;}} (main3);
		\end{scope}
		\end{tikzpicture}
	}
	\caption{State of the \textsf{stable} set after preprocessing according to \cite{Seidl2020}. Unknowns in \textsf{stable} are marked in green, unknowns that were removed from stable or are fresh are displayed in orange.
	For better visualization, unknowns referring to program locations are displayed within their corresponding control-flow-graph.
	}
	\label{fig:incr_destab}
\end{figure}
\end{example}

\paragraph{Reluctant destabilization} Immediate destabilization of some unknown $x$ may be considered unnecessary when the value of $x$ after the modification (i.e., with respect to the new system of equations) has not changed.
In particular, immediately destabilizing the unknowns corresponding to return nodes of function could be deemed inefficient when the state for the unknown would be the same after reanalysis of the function with the new constraint system.
In case the abstract value of $x$ has decreased in the lattice, we still may consider destabilization since the improved value for $x$ is expected to
reduce formerly flagged warnings.

To allow for such selective destabilization (and accordingly, selective reanalysis),
we propose a set $A$ of unknowns is chosen where the new value determines the need for further destabilization.
We proceed in two steps, where the solver is called for:
\begin{enumerate}[label=\arabic*.,ref=\arabic*]
	\item\label{item:unknown_a} all unknowns from $A$, and then
	\item\label{item:initial_unknown} the initial queried unknown.
\end{enumerate}
Any necessary destabilization for step \ref{item:unknown_a} is done by the fixpoint engine $\textbf{TD}_{\emph{side}}$, as is, which itself calls
\textsf{destabilize} for unknowns $x$ whenever their value is changed.
%
An important issue for reluctant destabilization is to identify the set $A$ of unknowns for the reanalysis of step~\ref{item:unknown_a}.
One choice is to consider reanalysis at the granularity of function definitions.
Then the set $A$ consists of all pairs $\angl{r,\alpha}$
where $r$ is the endpoint of some modified function $f$ and $\alpha$ is a calling-context of $f$,
while all other program points of the modified function are considered as \emph{new}, i.e.,
receive fresh identifiers which thus do not occur in any solver data structure so far.
When function headers are modified
(e.g., by adding another parameter) reluctant destabilization may perform work in vain and therefore is not applied to these functions.

\begin{example}\label{e:side_reluctant}
Again, consider the program from \cref{f:side}, with the modification of the assignment within
\textsf{foo} as presented in \cref{e:side_non_reluctant}.
With function granularity we have $A = \{\angl{2, \beta_0}\}$.
Nodes of the function are destabilized, but only until $A$, as shown in \cref{fig:reluctant_destab}.

According to step~\labelcref{item:unknown_a},
the unknown $\angl{2,\{p\mapsto\{\&g\}\}}$ is solved
starting from the previously attained solution $\rho$ from \cref{e:side2}.
During this, the modified assignment now provides an additional side-effect $\{2\}$ onto the global $g$,
which subsequently triggers a destabilization of the unknown
$\angl{5,\emptyset}$ corresponding to the endpoint of \textsf{main}.
The value for the unknown $\angl{2,\{p\mapsto\{\&g\}\}}$, on the other hand, does not change, therefore destabilization is not triggered there and the dependent
$\angl{4,\emptyset}$ remains in \textsf{stable}, as shown in \cref{fig:reluctant_destab2}.
Now according to step~\labelcref{item:initial_unknown},
the unknown $\angl{5,\emptyset}$ is solved.
Only a single right-hand side evaluation is needed to find the new value $\{\textsf{ret}\mapsto\{0,1,2\}\}$ of this unknown before reanalysis terminates.\qed

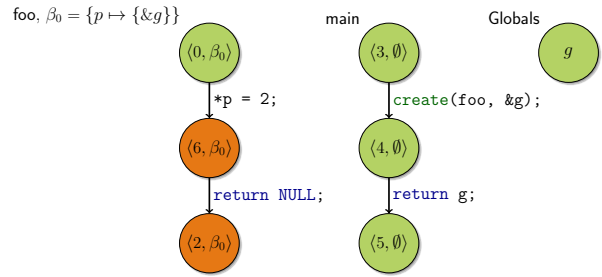
\begin{figure}
	\resizebox{0.45\textwidth}{!}{
		\begin{tikzpicture}[auto,
			node distance = 12mm,
			start chain = going below,
			f/.style = {font = \LARGE},
			ed/.style = {f, ultra thick},
			state/.style = {draw,circle,blur shadow,fill=norange,
					on chain,align=center, minimum size=2cm, f},
			stable/.style = {fill=lgreen}]

		\node[state, stable] (foo1)    {$\angl{0,\beta_0}$};
		\node[state] (foo2)    {$\angl{6,\beta_0}$};
		\node[state] (foo3)    {$\angl{2,\beta_0}$};
		\node (foo_label)[above left=0.2cm of foo1, f] {\textsf{foo}, $\beta_0=\{p\mapsto\{\&g\}\}$};

		\node[state,stable] (main1)[right=4cm of foo1]    {$\angl{3, \emptyset}$};
		\node[state,stable] (main2)    {$\angl{4, \emptyset}$};
		\node[state,stable] (main3)    {$\angl{5, \emptyset}$};
		\node (main_label)[above left=0.2cm of main1, f] {\textsf{main}};

		\node[state, stable] (g)[right=4cm of main1] {$g$};
		\node (global_label)[above left=0.2cm of g, f] {\textsf{Globals}};

		\begin{scope}[rounded corners]
		\path [->] (foo1) edge[ed] node {\lstinline{*p = 2;}} (foo2);
		\path [->] (foo2) edge[ed] node {\lstinline{return NULL;}} (foo3);

		\path [->] (main1) edge[ed] node {\lstinline{create(foo, \&g);}} (main2);
		\path [->] (main2) edge[ed] node {\lstinline{return g;}} (main3);
		\end{scope}
		\end{tikzpicture}
}
\caption{State of the \textsf{stable} set after preprocessing with reluctant destabilization.
Unknowns relating to nodes of \textsf{foo} after the entry-node are removed from \textsf{stable}, but the unknown $\angl{2, \beta_0} \in A$ is not destabilized.}
\label{fig:reluctant_destab}
\end{figure}

\begin{figure}
	\resizebox{0.45\textwidth}{!}{
		\begin{tikzpicture}[auto,
			node distance = 12mm,
			start chain = going below,
			f/.style = {font = \LARGE},
			ed/.style = {f, ultra thick},
			state/.style = {draw,circle,blur shadow,fill=norange,
					on chain,align=center, minimum size=2cm, f},
			stable/.style = {fill=lgreen}]

		\node[state,stable] (foo1)    {$\angl{0,\beta_0}$};
		\node[state,stable] (foo2)    {$\angl{6,\beta_0}$};
		\node[state,stable] (foo3)    {$\angl{2,\beta_0}$};
		\node (foo_label)[above left=0.2cm of foo1, f] {\textsf{foo}, $\beta_0=\{p\mapsto\{\&g\}\}$};

		\node[state,stable] (main1)[right=4cm of foo1]    {$\angl{3, \emptyset}$};
		\node[state,stable] (main2)    {$\angl{4, \emptyset}$};
		\node[state] (main3)    {$\angl{5, \emptyset}$};
		\node (main_label)[above left=0.2cm of main1, f] {\textsf{main}};

		\node[state, stable] (g)[right=4cm of main1] {$g$};
		\node (global_label)[above left=0.2cm of g, f] {\textsf{Globals}};

		\begin{scope}[rounded corners]
		\path [->] (foo1) edge[ed] node {\lstinline{*p = 2;}} (foo2);
		\path [->] (foo2) edge[ed] node {\lstinline{return NULL;}} (foo3);

		\path [->] (main1) edge[ed] node {\lstinline{create(foo, \&g);}} (main2);
		\path [->] (main2) edge[ed] node {\lstinline{return g;}} (main3);
		\end{scope}
		\end{tikzpicture}
}
\caption{State of the \textsf{stable} set after reluctant destabilization solves $\angl{2, \beta_0} \in A$ during step~\labelcref{item:unknown_a}.}
\label{fig:reluctant_destab2}
\end{figure}
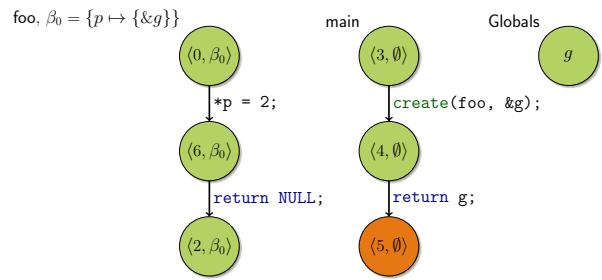
\end{example}

To maximize reuse, we would like to identify as many program points as possible whose
corresponding unknowns have unchanged right-hand sides.
A finer granularity can be obtained by means of a (possibly partial) \emph{matching} between the control-flow graphs of the two versions
of the modified functions.
We build on the \textsc{diffcond} algorithm from \cite{BeyerJL20} that
obtains a weakly connected sub-graph of the cfg that is unchanged between two versions of a function.
We add a phase to the algorithm that removes all nodes from the matching, for which any of the predecessors is changed or new.
This way, we can ensure that the matched part of the graph only contains nodes,
for whose corresponding unknowns the right-hand sides have not changed.

%
For each modified function $f$, the matching provides us with a partial injective mapping $\eta_f$ from the program points of
the original implementation of $f$ to the program points of the updated version.
Thereby, we demand that endpoints are preserved.
The mapping $\eta_f$ is used to transfer identities of program points across the update: only the program points not occurring in the
image of $\eta_f$, are now considered as \emph{new}.
The set $A$ of unknowns scheduled for reanalysis in the first stage, still consists of the set of all pairs $\angl{r,\alpha}$
where $r$ is the endpoint of some modified function $f$ and $\alpha$ is a calling-context of $f$.
The unknowns $\angl{\eta_f\,v,\alpha}$ correspond to program points which were already present in the control-flow
graph of the original version of $f$, and therefore are considered as \emph{not new}.
Of these, we remove every $\angl{\eta_f\,v,\alpha}$ from \textsf{stable} where the list of incoming edges of $\eta\,v$
differs from the list of incoming edges of $v$ in the original control-flow graph of $f$
-- together with all unknowns $\angl{v',\alpha}$ where $v'$ is reachable from $\eta_f\,v$ in the
control-flow graph for the new version of $f$.

\begin{example}\label{e:side_reluctant_cfg}
Consider the program example of \cref{f:side} from the last section,
and assume that the function \textsf{foo} has been modified so that now the first assignment in its body is changed to
\begin{lstlisting}
*p = 2;
\end{lstlisting}
while everything else stays the same.
Accordingly, we may \emph{re-use} the program points of the original version also for the new version of \textsf{foo}.
Then before re-evaluation at stage 1,
the unknown $\angl{1,\{p\mapsto\{\&g\}\}}$ is removed from the set \textsf{stable} since the label of its incoming edge
has been changed. Accordingly, also the unknown $\angl{2,\{p\mapsto\{\&g\}\}}$ corresponding to the endpoint of \textsf{foo}
in calling-context $\{p\mapsto\{\&g\}\}$ is removed from \textsf{stable}.

Reanalysis
of the unknown $\angl{2,\{p\mapsto\{\&g\}\}}$
starting from the previously attained solution $\rho$ from \cref{e:side2},
then provides an additional side-effect $\{2\}$ onto the global $g$,
which subsequently triggers a destabilization of the unknown
$\angl{5,\emptyset}$ corresponding to the endpoint of \textsf{main}.
The value for the unknown $\angl{2,\{p\mapsto\{\&g\}\}}$, on the other hand, does not change, so that
$\angl{4,\emptyset}$ remains in \textsf{stable}.

Reanalysis of the unknown $\angl{5,\emptyset}$ at stage 2 therefore only re-evaluates the right-hand side of this unknown
to find its new value $\{\textsf{ret}\mapsto\{0,1,2\}\}$ before reanalysis terminates.
\qed
\end{example}
\end{comment}

%% file: content/04-restart.tex
\section{Restarting Affected Globals \label{sec:restart}}

Analyzing globals flow- and context-insensitively (as in \cref{e:side}), is not only helpful
for thread-modular static analysis, but also for increasing incrementality.
As observed in \cite{Seidl2020}, already minor changes of globals may trigger whole program
reanalysis, when globals are analyzed flow- and context-sensitively, i.e., as part of the local state.
The downside of analyzing globals flow- and context-insensitively, however, is that the abstract values for these
are accumulated over multiple fixpoint iterations and thus can only become larger
(with regard to the order of the domain lattice), i.e., more \emph{imprecise}.

\begin{example}\label{e:side4}
In \cref{e:side_reluctant}, we have considered the modification of
the program from \cref{f:side} by changing
the first line of the body of function \textsf{foo} to
{\small
\begin{lstlisting}
*p = 2;
\end{lstlisting}}
%
Reanalysis of the system according to the methods from the last section,
when starting from the previously attained solution $\rho$ from \cref{e:side2}
additionally takes the new side-effect $\{2\}$ for $g$ into account
and increases the set of possible values of $g$ from $\{0,1\}$ to $\{0,1,2\}$.
A \emph{from-scratch} analysis of the modified program would instead find
the set $\{0,2\}$ of possible values of $g$.
\qed
\end{example}
\paragraph{Restarting flow-insensitive unknowns}

Let us call unknowns which receive all their values via side-effects, \emph{flow-insensitive}.
If the value of a flow-insensitive unknown $g$ before reanalysis is involved in a warning,
the incremental reanalysis of the modified code may only increase the value of $g$ implying that its result cannot exclude the warning
even if the issue has been fixed.
As one way to remedy this loss of precision, we propose to \textit{restart} the evaluation of (selected) flow-insensitive unknowns.
Restarting a subset $G$ of flow-insensitive unknowns means that for every $g\in G$,
the abstract value is set to $\bot$ and \textsf{destabilize} $g$ is called.
Furthermore, care has to be taken that \emph{all} side-effects to unknowns in $G$ of the modified the equation system
will be re-evaluated during reanalysis.
This is achieved by removing the unknowns $x\in\textsf{side\_dep}\,g, g\in G$, from the set \textsf{stable}
and calling \textsf{destabilize} for each of them as well.
A \emph{minimal} strategy sticks with this.

\begin{example}\label[example]{e:side5}
Continuing with the \cref{e:side4},
we may want to restart the set $G=\{g\}$.
Side-effects to $g$ occur inside the global initializer \textsf{init} (within the body of the harness \textsf{\_\_main})
as well as at the replaced $\angl{1,\{p\mapsto\{\&g\}\}}$ inside the function \textsf{foo}.
Destabilization of the corresponding unknowns removes from the set \textsf{stable} the unknowns
$\angl{4,\emptyset}, \angl{5,\emptyset}$ as well as
$\angl{1,\{p\mapsto\{\&g\}\}},\linebreak[1] \angl{2,\{p\mapsto\{\&g\}\}}$.
The state of the preprocessed data after this restarting is shown in \cref{fig:restarting}.
Running the solver on the such preprocessed data structures, results in an assignment which maps $g$ to the set $\{0,2\}$ and
$\angl{5,\emptyset}$ to the mapping $\{\textsf{ret}\mapsto\{0,2\}\}$.
Indeed, the artifact value 1 is purged from all reported value sets.

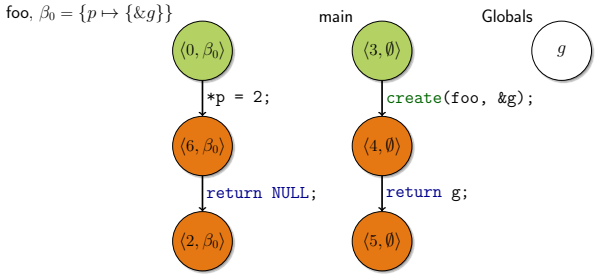
\begin{figure}
	\resizebox{0.45\textwidth}{!}{
		\begin{tikzpicture}[auto,
			node distance = 12mm,
			start chain = going below,
			f/.style = {font = \LARGE},
			ed/.style = {f, ultra thick},
			state/.style = {draw,circle,blur shadow,fill=norange,
					on chain,align=center, minimum size=2cm, f},
			stable/.style = {fill=lgreen},
      reset/.style = {fill=white}]

		\node[state, stable] (foo1)    {$\angl{0,\beta_0}$};
		\node[state] (foo2)    {$\angl{6,\beta_0}$};
		\node[state] (foo3)    {$\angl{2,\beta_0}$};
		\node (foo_label)[above left=0.2cm of foo1, f] {\textsf{foo}, $\beta_0=\{p\mapsto\{\&g\}\}$};

		\node[state,stable] (main1)[right=4cm of foo1]    {$\angl{3, \emptyset}$};
		\node[state] (main2)    {$\angl{4, \emptyset}$};
		\node[state] (main3)    {$\angl{5, \emptyset}$};
		\node (main_label)[above left=0.2cm of main1, f] {\textsf{main}};

		\node[state, reset] (g)[right=4cm of main1] {$g$};
		\node (global_label)[above left=0.2cm of g, f] {\textsf{Globals}};

		\begin{scope}[rounded corners]
		\path [->] (foo1) edge[ed] node {\lstinline{*p = 2;}} (foo2);
		\path [->] (foo2) edge[ed] node {\lstinline{return NULL;}} (foo3);

		\path [->] (main1) edge[ed] node {\lstinline{create(foo, \&g);}} (main2);
		\path [->] (main2) edge[ed] node {\lstinline{return g;}} (main3);
		\end{scope}
		\end{tikzpicture}
  }
  \caption{State of the analysis data after restarting $G=\{g\}$, but before reanalysis. Unknowns in \textsf{stable} are displayed in green.
  The unknown $g$ was reset to $\bot$ (displayed in white).}
  \label{fig:restarting}
\end{figure}
%
Using \emph{reluctant} destabilization, i.e., only up to the respective return nodes of functions,
only the nodes
$\angl{1,\{p\mapsto\{\&g\}\}}, \angl{2,\{p\mapsto\{\&g\}\}}$ and $\angl{5,\emptyset}$ are removed
from \textsf{stable} -- resulting in the same analysis result with less effort.
\qed
\end{example}

\paragraph{Selecting unknowns for restarting}
Instead of explicitly  specifying the set $G$ of global unknowns to be restarted,
we may extract for a modified function $f$, the set $G_f$ of all globals to which side-effects occurred
in the right-hands of unknowns for the old version of $f$ -- which may no longer occur
in the right-hand sides of unknowns for the current version.
For that, the mapping \textsf{side\_infl} is consulted for every unknown $\angl{u,\alpha}$, $u$ being a program point of the old version of $f$
and $\alpha$ being a calling-context.

The \emph{minimal} strategy does not take into account that too imprecise values for the unknowns in $G$
may have an impact also onto the side-effects triggered for \emph{further} globals.
A more ambitious and, accordingly, also more expensive strategy tries to purge the analysis result also of these indirect
losses of precision.
%
According to our preliminary experiments, the price to be paid for that, though, is a significant increase in interactive reanalysis time.
Therefore, we do not pursue this option further.
For that, we introduce the function \textsf{destabilize\_with\_side}, shown in \cref{l:destab_side} to be applied to all unknowns in $G$.
\begin{lstlisting}[language={[Objective]Caml}, gobble=2, label=l:destab_side, float, caption={Function destabilize with side.}]
  let rec destabilize_with_side p x =
    let destab x m =
      let w = m x in
      m x := |$\emptyset$|;
      iter (fun y ->
        stable -= y;
        destabilize_with_side p y
        ) w
    in
    destab x infl;
    destab x side_infl;
    if p x then (
      |$\rho$| x := |$\bot$|;
      destab x side_dep;
    )
\end{lstlisting}
The function \textsf{destabilize\_with\_side} receives as an argument a predicate \textsf{p} which indicates which unknowns are candidates for
restarting. In the most ambitious setting, which we might call \emph{exhaustive}, we set \textsf{p x = true}.
More refined definitions are conceivable, though, in order to gradually trade efficiency for precision, e.g., cutoff by recursion depth.
%
%
The \textsf{destabilize\_with\_side}  relies on the helper function \textsf{destab} which takes an unknown \textsf{x} and
a map \textsf{m} providing for each unknown \textsf{y} another set \textsf{m y} of unknowns with which destabilization should proceed.
The helper function first stores the value of \textsf{x} in \textsf{m} in some local variable \textsf{w}
before resetting the value of \textsf{m} for \textsf{x} to the empty set.
Each unknown \textsf{y} in \textsf{w} then is removed from the set \textsf{stable}
before \textsf{destabilize\_with\_side p} is recursively called for \textsf{y}.
The call \textsf{destabilize\_with\_side p x} then calls \textsf{destab x} first for the map \textsf{infl} and then for the map \textsf{side\_infl}.
If additionally \textsf{p x} holds,
the value $\rho\,\textsf{x}$ is reset to $\bot$, and
\textsf{destab x side\_dep} is called.

The set $G$ of initially selected unknowns for restarting may either be selected by the user,
or can automatically be inferred as the set of globals receiving side-effects from unknowns that have been removed from the
set \textsf{stable} before reanalysis.

\begin{example}\label{e:sideside}
\begin{figure}
\begin{minipage}{0.24\textwidth}\small
\begin{lstlisting}[basicstyle=\footnotesize]
atomic int g = 0 ;
atomic int h = 0 ;
void* foo(void* p) {
   h = 1;
   *p = h;
   return NULL;
}
int main() {
   create(foo, &g);
   return g;
}
\end{lstlisting}
\end{minipage}%
\begin{minipage}{0.24\textwidth}
\resizebox{1\textwidth}{!}{
\begin{tikzpicture}[auto,
   node distance = 12mm,
   start chain = going below,
   state/.style = {draw,circle,blur shadow,fill=white,
         on chain,align=center}
         ]
  \node[state] (foo1)    {$0$};
  \node[state] (foo2)    {$1$};
  \node[state] (foo6)    {$6$};
  \node[state] (foo3)    {$2$};
  \node (foo_label)[above left=0.2cm of foo1] {\textsf{foo}};

  \node[state] (main1)[right=2.5cm of foo1]    {$3$};
  \node[state] (main2)    {$4$};
  \node[state] (main3)    {$5$};
  \node (main_label)[above left=0.2cm of main1] {\textsf{main}};

  \begin{scope}[rounded corners]
   \path [->] (foo1) edge node {\lstinline{h = 1;}} (foo2);
   \path [->] (foo2) edge node {\lstinline{*p = h;}} (foo6);
   \path [->] (foo6) edge node {\lstinline{return NULL;}} (foo3);

   \path [->] (main1) edge node {\lstinline{create(foo, \&g);}} (main2);
   \path [->] (main2) edge node {\lstinline{return g;}} (main3);
  \end{scope}
 \end{tikzpicture}
}
\end{minipage}
\caption{\label{f:sideside}A multithreaded code with side-effects onto two globals.}
\end{figure}
In the example program in \cref{f:sideside}, the side-effect onto the global $h$ is apparent from the code of the function \textsf{foo}, while the side-effect onto
$g$ only occurs due to the given calling-context $\{p\mapsto\{\&g\}\}$.
After replacement of the assignment to the global \textsf{h} in the body of \textsf{foo} to
\begin{lstlisting}
h = 2;
\end{lstlisting}
the set $G_{\textsf{foo}}$ is found to be $\{g,h\}$ -- implying that before reanalysis, not only the value of $h$,
but also the value of $g$ is reset to $\bot$.
\qed
\end{example}
\end{comment}

\paragraph{Write-only globals}
In some cases, flow-insensitive unknowns are introduced in order to collect information which does not affect the analysis itself, but is later used
in the postprocessing phase for the generation of warnings. In particular, this is the case for the unknowns recording accesses to globals or lock operations as used by the \emph{lockset} and \emph{lock-order} analyses as described in \cref{sec:implementation}.
Technically, such a \emph{write-only} unknown $\textit{acc}$ collects values into a \emph{set} domain
with the following two properties:
\begin{enumerate}[label=(WO\arabic*), align=left]
  \item\label{i:unique-trace-back} \emph{unique trace-back}: each contribution can uniquely be traced back to the unknown producing it, and
  \item\label{i:purely-accumulating} \emph{purely accumulating}: the unknown must never be read during the analysis.
\end{enumerate}
When restarting $\textit{acc}$,
the analysis need not reset the value of $\textit{acc}$ to $\bot$, but may decide for each unknown $x$ with a contribution to the value of $\textit{acc}$ whether to
\begin{enumerate}[label=(\alph*)]
\item	remove the contribution, but destabilize $x$, or
\item	retain the contribution and forgo destabilization.
\end{enumerate}
Instead of \emph{restarting} the unknown $\textit{acc}$, however,
the computation of the value of $\textit{acc}$ during the fixpoint iteration itself can be completely abandoned (as it is never read)
by deferring it to
the postprocessing phase. This is made precise in the next section.
%

%

%% file: content/05-pipeline.tex
\section{Interactive Analysis Pipeline}
\label{sec:pipeline}
\label{sec:implementation}
An overview of the structure of the interactive analyzer is shown in \cref{f:pipeline}.
On top of the changes to solving, further improvements to the entire analysis pipeline are needed in order to make the whole analysis \emph{incremental} and fast.
In particular, this applies to the \emph{postprocessing} to generate warnings from the analysis results.

\paragraph{Postprocessing.}
Once the solver has terminated, the \linebreak[4]equation system solution is postprocessed:
\begin{enumerate}
    \item Auxiliary data for the incremental solver (\textsf{side\_infl}, \textsf{side\_dep}) is collected;
    \item Warnings are generated;
    \item The solution is pruned of unreachable unknowns.
\end{enumerate}
In ~\cite{Seidl2020}, e.g., this was done by an additional \emph{full} re-evaluation pass over the solution.
This, however, is expensive and redundant when changes to the program, the solution, and the warnings are relatively small.
\linebreak[4]Therefore, we also incrementalize this process.
We introduce an extra set \textsf{superstable} consisting of unknowns which were in the set \textsf{stable} before reanalysis and stayed therein throughout.
This implies that their abstract values are unchanged, and the warnings generated from them can be reused.
The postprocessing re-evaluations are then limited to unknowns \emph{not} in \textsf{superstable}.

Only warnings from added or changed unknowns are generated because those unknowns were not in \textsf{superstable} to begin with or were at some point removed by the solver's destabilization, respectively.
Since warnings are associated with CFG nodes and equation system unknowns, reused warnings still get updated source code locations when the corresponding code has moved due to surrounding changes, irrelevant whitespace changes, etc.

Reluctant destabilization from \cref{sec:reluctant} can
reduce the number of unknowns removed from \textsf{stable}.
In the setting with incremental postprocessing, this in turn means
that fewer unknowns may be removed from \textsf{superstable}, which additionally reduces the workload for the incremental postprocessing.
In case of data-races as well as deadlocks, warnings are generated from \emph{write-only} unknowns satisfying the properties
\labelcref{i:unique-trace-back} and \labelcref{i:purely-accumulating} from the previous section.
As another optimization,
side-effects to these unknowns are no longer executed during the (re-)analysis itself - but only emitted during the postprocessing re-evaluations of
right-hand sides, while the side-effects from unknowns from the set \textsf{superstable} are reused.
By this, expensive restarting from \cref{sec:restart} is avoided.
%

%% file: content/06-practical.tex

\paragraph{Analyses}
For concurrent programs, we consider
\begin{itemize}
\item	a \emph{lockset} analysis in order to verify absence of data-races
	(see, e.g., \cite{VojdaniVene09,Goblint2016} for background and further related work);
\item	a \emph{lockorder} analysis \cite{Deadlock/KroeningPSW16} to detect potential deadlocks.
\end{itemize}
%
The lockset analysis maintains for each program point the set of (addresses of) definitely held mutexes.
%
%
For each abstract possibly shared data element $g$, the set of \emph{accesses} $\textsf{acc}_g$ to it is collected,
recording its program location,
whether the access was a read or write and the set of definitely held mutexes.
%
%
The sets of accesses then are used for generating data-race warnings.

The lockorder analysis introduces for each mutex $A$, a flow-insensitive unknown $\textsf{ord}_A$ which collects
\emph{locking events} that may occur while the mutex $A$ is held.
Such a locking event consists of the program point, where the lock operation is executed, together with the mutexes which have possibly been acquired.
The values of the unknowns $\textsf{ord}_A$ are used to detect potential circular locking of
mutexes. Again, \emph{may-happen-in-parallel} information may be used to reduce the number of false alarms.

After a flagged data-race for the shared data element $g$ has (potentially) been fixed,
the unknown $\textsf{acc}_g$ must be purged of former accesses.
Otherwise, race warnings would accumulate over sequences of reanalyses, and never be removed.
A similar effect occurs for deadlock warnings.
%
Purging of these write-only unknowns can be achieved efficiently using the technique described above.

\begin{figure}[t]
    \includegraphics[width=0.45\textwidth]{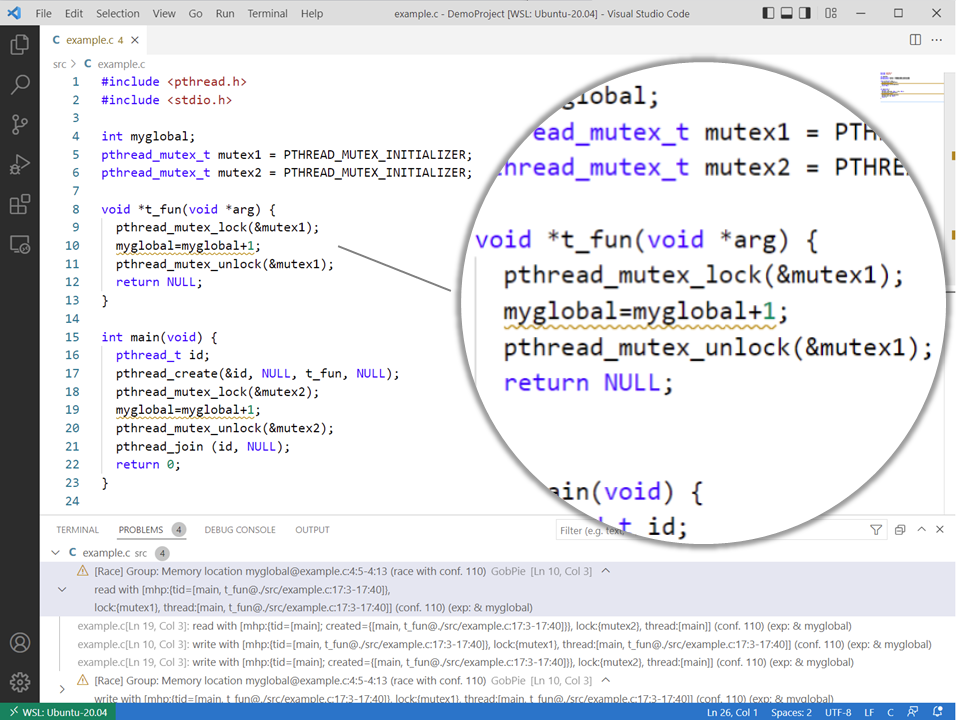}
    \caption{\textsc{Goblint} integration via \textsc{MagpieBridge} displaying a race warning in \textsc{Visual Studio Code}.}
    \label{f:gobpie}
\end{figure}

\paragraph{IDE integration}
The \textsc{MagpieBridge} framework \cite{luo_et_al:LIPIcs:2019:10813} aims to provide a generic communication layer between static analysis tools and various Integrated Development Environments.
The framework simplifies integration as only one binding has to be implemented for a static analysis tool to support a range of IDEs.
Our binding communicates with \textsc{Goblint} (running in server mode) over a Unix socket using JSON-RPC to trigger interactive reanalysis on code changes. Analysis results from \textsc{Goblint} are converted and forwarded for display in the IDE, making the analysis automatic and seamless.
\Cref{f:gobpie} shows the integration in action.
The integration is available in the artifact~\cite{artifact}.

%% file: content/07-consistency.tex
\section{Experimental Evaluation}
\label{sec:evaluation}
\label{sec:questions}

In order to evaluate the potential of interactive abstract interpretation in a multithreaded setting,
we conducted extensive experiments in order to answer to the following research questions:

\begin{description}
\item[RQ1] \emph{Efficiency:} Do our techniques effectively reduce the amount of reanalysis after a code change?
\item[RQ2] \emph{Consistency:} What is the price to be paid in terms of precision for more efficient reanalysis?
\item[RQ3] \emph{Usability:} Can the system be conveniently used for the development of concurrent code?
\end{description}
%

%% file: content/08-evaluation.tex
\begin{figure*}[h]
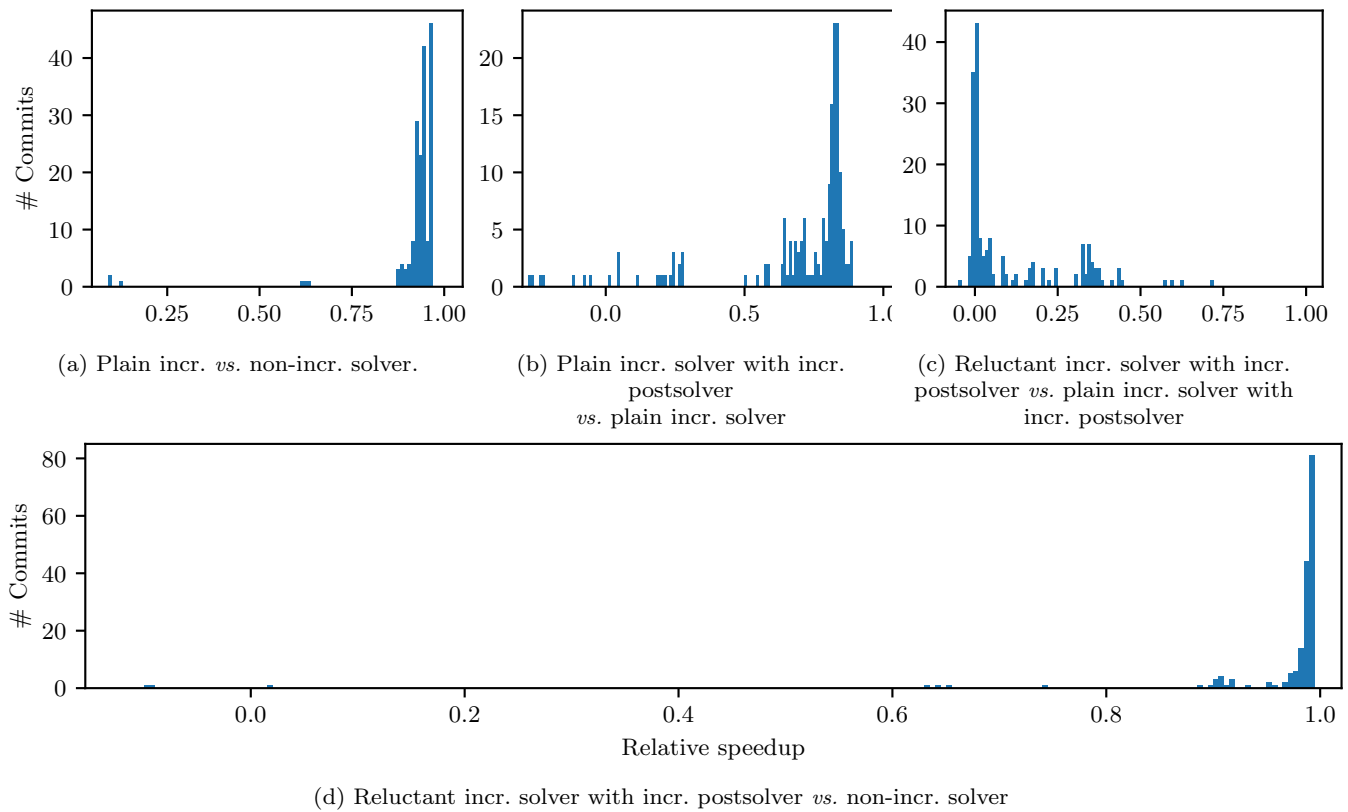

  \captionsetup[subfigure]{justification=centering}
  \begin{center}
    \begin{subfigure}[t]{0.35\textwidth}
      \centering
      \input{figures/efficiency_figure_0.pgf}
      \subcaption{Plain incr. \emph{vs.} non-incr.\ solver. }
      \label{fig:12}
    \end{subfigure}%
    \begin{subfigure}[t]{0.32\textwidth}
      \centering
      \input{figures/efficiency_figure_1.pgf}
      \subcaption{Plain incr. solver with incr. postsolver\\ \emph{vs.} plain incr. solver}
      \label{fig:13}
    \end{subfigure}%
    \begin{subfigure}[t]{0.32\textwidth}
      \centering
      \input{figures/efficiency_figure_2.pgf}
      \subcaption{Reluctant incr. solver with incr. postsolver \emph{vs.} plain incr. solver with incr. postsolver}
      \label{fig:14}
    \end{subfigure}
    \bigskip
    \begin{subfigure}[b]{\textwidth}
      \input{figures/efficiency_figure_3.pgf}
      \subcaption{Reluctant incr. solver with incr. postsolver \emph{vs.} non-incr. solver}
      \label{fig:11}
    \end{subfigure}%
  \end{center}
  \label[figure]{figs:efficiency}
  \caption{\cref{fig:12,fig:13,fig:14} show the number of commits for which a given relative speedup for reanalysis setups (1)-(3)
  w.r.t.\ the respective baseline setting was achieved.
  (1) is compared to the \emph{from-scratch} analysis, while (2) and (3) are compared to (1) and (2) respectively.
  \cref{fig:11} highlights the relative overall speedup by comparing (3) with the \emph{from-scratch} analysis.}
\end{figure*}
As our benchmark suite, we considered the commits from the open-source \textsc{Git} repository for
\textsc{zstd}\footnote{\url{https://www.github.com/facebook/zstd}}.
The system implements a compression algorithm and relies on a set of worker threads to exploit the availability of multiple cores.
We extract commits using
\textsc{PyDriller}~\cite{PyDriller}.
In the current version, the program to be analyzed consists of around 22,000 lines of C code (excluding comments).
It has been developed at least since 2015 with about 9,500 commits. From these, we concentrated on the commits from Aug. 1, 2021 to Jan. 31, 2022.
The same analysis specification is used for all experiments.
Besides, the lockset analysis, it consists of
\begin{enumerate}
    \item a base value analysis 
	  for arbitrary C-data types, including
exclusion-set domains for integers (for more precise treatment of switch statements),
a points-to analysis for pointers and
pointer must-equalities~\cite{Seidl09}.
    \item
\emph{symbolic} locks
in order to deal with dynamically allocated data hosting their individual mutexes \cite{Goblint2016}.
    \item a plain thread analysis which distinguishes between the \emph{main} thread and started threads and also between single-threaded and
multithreaded mode.
\end{enumerate}
Experiments for \textbf{RQ1} and \textbf{RQ2} were executed on a machine with two Intel Xeon Platinum 8260 CPUs, each with 48 physical cores at 2.4GHz (Turbo Boost disabled), and
256 GB RAM under Ubuntu 18.04.
While the analyzer \textsc{Goblint} runs on a single core, the host's concurrency was used to parallelize the analysis of different commits.
In order to obtain reliable results, we followed the suggestions in \cite{Beyer2019}:
We pinned individual analysis tasks to individual cores to get reproducible runtimes and tried to distribute work so that
an equal number of tasks shore the same L3 cache (of which the machine provides only two).
Also, we only used 20 of the available cores in order to reduce the variance of runtime measurements.

As measures of performance, we report total analyzer runtime.
It includes the wall time taken by all steps of the analysis pipeline, including parsing, control-flow-generation,
solving of the constraint system, and outputting warnings.
In non-incremental analyses, solving the constraint system is responsible for the dominating share of wall clock time.
In turn, the evaluation of right-hand sides makes up a significant portion of solving time.

For answering \textbf{RQ1} (efficiency), we approximated
the single-step modification of a developer by suitable commits from the \textsc{zstd} repository.
We only considered not too large commits (at most 50 changed lines in C or header files) with a single parent --
of which we find 322 out of which only 183 resulted in changes to the C code to be analyzed.
Subsequently, we report numbers only for these.
Here, \emph{from-scratch} analysis times range from about 3000 to 5000s.
%
In the first three runs, we compare the \emph{from-scratch} analysis time of the parent version to the reanalysis time for the given commit.
For reanalysis, we consider the following setups (see \cref{fig:12,fig:13,fig:14}):
\begin{enumerate}
\item	The plain incremental solver from \cite{Seidl2020} enhanced with minimal restarting for the auxiliary globals
	$\textsf{acc}_g$ collecting the accesses to global variables $g$;
\item	The plain incremental solver from \cite{Seidl2020} together with the incremental postprocessing from \cref{sec:pipeline};
\item	The reluctant solver from \cref{sec:reluctant} together with the incremental postprocessing from \cref{sec:pipeline}.
\end{enumerate}
A \emph{relative} improvement of $0$ indicates that runtime remained the same, whereas a relative improvement of $1$ would indicate
that running in the second setting takes no time at all.

In summary, we obtain that for more than 80\% of the commits, the incremental reanalysis with restarting only requires less than
8\% of the \emph{from-scratch} analysis time of the parent commit.
%
%
Incremental postprocessing resulted in another boost in efficiency by
reducing the overall time for reanalysis in more than 80\% of all cases to about 3\% of the \emph{from-scratch} analysis time.
A single outlier is omitted from the plot, where incremental postprocessing was slower than non-incremental postprocessing by a factor of about 3
(but still faster than from-scratch analysis by a factor of about 3).
The impact of reluctant destabilization is less clear.
While reluctant destabilization does not impact the efficiency in all cases, a further reduction to about 2\% of the
from-scratch runtime can be observed for at least 75\% of the commits.
%

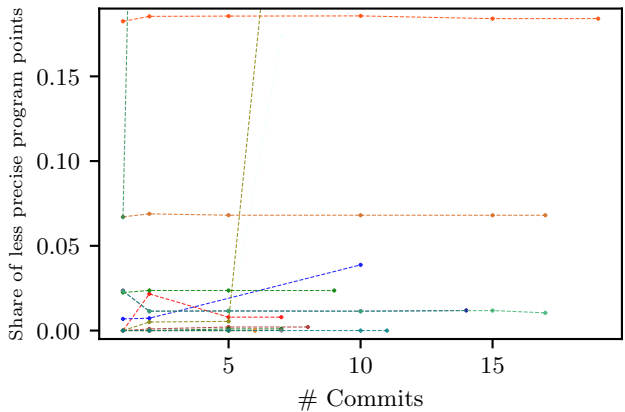
\begin{figure}
  \centering
  \input{figures/precision_figure.pgf}
  \caption{Share of program points less precise than in a \emph{from-scratch} run after a sequence of $x$ commits.}
  \label{fig:deter}
\end{figure}

Concerning \textbf{RQ2} (consistency),
let us call the result of a reanalysis for the modified program \emph{from-scratch consistent}
if it is less or equal (w.r.t.\ the ordering of the abstract domains)
to the analysis result provided by a from-scratch analysis of the modified program \cite{Stein21}.
There are several reasons from-scratch-consistency in all cases is out of reach.
These include \emph{non-monotonicity} of right-hand sides in a context-sensitive setting;
use of \emph{widening} and \emph{narrowing} to cope with infinite ascending/descending chains in domains and
the accumulative treatment of flow-insensitive unknowns.
Therefore, we track the deterioration of precision across series of reanalyses.
We consider only the most efficient configuration (3).
In our benchmark repository, we singled out sequences starting with a merge commit, followed by at least five single parent commits.
In our time frame, we found 18 such sequences with an average length around 10.
We excluded one sequence that produced parse errors.
Two sequences of length 14 showed extraordinary losses in precision, namely, were less precise on more than 60\% of program points after
two and at least five incremental reanalyses, respectively.
These are only included into \cref{fig:deter} up to the reanalysis before which the dramatic loss occurs.
%
For each remaining sequence, we compared the result of \emph{from-scratch} analysis after 1, 2, 5, 10, and 15 incremental commits
and of the respective version with the result obtained by repeated reanalysis.
As a measure of consistency of analysis results, we relied on the number of program points where reanalysis results are less precise
than from-scratch analysis results.
The results (see \cref{fig:deter})
indicate that for all sequences,
virtually no loss occurs (throughout all versions at most 5\% of program points were affected).
Only one sequence (in addition to the two discussed above) exhibited a higher loss
at between 15 and 20\%.
%
%

\paragraph{Threats to validity.}
Our experimental evaluation for \textbf{RQ1} and \textbf{RQ2} relies on the commits to a single repository over six months only.
This project, however, is an industry-grade project backed by a large company where the employed development pattern can be expected
to be prototypical for similar long-running projects.
It is still debatable whether small commits to a \textsc{Git} repository can serve as realistic proxies for interactive editing sessions.
Best practices, on the one hand, encourage developers to commit small increments regularly.
On the other hand, using such real-world modifications is clearly preferable to artificially created editing sequences.
Concerning the runtime measurements for answering \textbf{RQ1},\linebreak[4]
there may be further variations in the exact runtimes due to caching or concurrency effects of our parallelized evaluation setup.
This would, however, affect both incremental and non-incremental analysis runs and even out across the large number of measurements.
Generally, we expect less significant speedups if the code base to be analyzed is small (see, e.g., die example programs
illustrating usability below).
Concerning \textbf{RQ2}, the comparison of analysis results between from-scratch analysis and repeated reanalysis is not
completely reliable, since the employed abstract domains do not fully agree:
after an incremental modification, \emph{may-points-to} sets may additionally contain
abstract heap locations created at program points which are no longer reachable. These render the incremental reanalysis less precise --
despite their presence being benign.
Accordingly, our measurements systematically \emph{underestimate} precision.
The analyzer, benchmarking scripts and raw results are available in the artifact~\cite{artifact}.

\medskip

Concerning \textbf{RQ3} (usability), we provide two slice-of-life editing scenarios: one in which a developer attempts to verify race freedom in a correct program, and one in which they attempt to fix race conditions in a buggy program.\footnote{Original source code and described patches are available in the artifact~\cite{artifact}.} While the latter is a fairly common editing scenario, the first is an approach advocated by Delmas and Souyris  \cite{Delmas07}:
a sound analyzer is used to certify the absence of run-time errors in real-world programs with the help of user annotations.
We evaluate how feasible this approach is with the \textsc{Goblint} analyzer and whether interactivity improves the experience.

\textsc{Chrony}\footnote{\url{https://chrony.tuxfamily.org/}} is an implementation of the Network Time Protocol (NTP) using multiple threads to resolve DNS queries asynchronously.
We analyze version~4.2
(consisting of 11606 logical LoC) and configure \textsc{Goblint} to use symbolic locksets and pointer must-equalities~\cite{Seidl09}, as well as freshness analysis for dynamic memory.
Acting as a developer, we go through the following workflow:
\begin{enumerate}
    \item Initial from-scratch analysis takes 4 minutes 42 seconds and finds two races.
    Manual inspection reveals that these are false alarms, due to \textsc{Goblint} not detecting some joined threads.
    The developer adds a \verb|__goblint_assume_join| annotation to guide the analyzer.
    \item Reanalysis after the change takes 26 seconds and finds no races --- the program is race-free!
    Manual inspection suggests that using a global mutex \verb|privops_lock| for protecting the worker thread body
    is not required in the program's default configuration.
    The developer removes the mutex to allow increased parallelism.
    \item Reanalysis after the change takes 14 seconds, but does not succeed in proving the absence of races.
    The developer implements more fine-grained synchronization by adding a mutex to the dynamically allocated struct that needs protection.
    \textsc{Goblint} can analyze such locking schemes by using \emph{symbolic} locksets.
    \item Reanalysis after the change takes 16 seconds and succeeds in proving the absence of races!
    The program is now without a single global mutex, allowing parallelism.
\end{enumerate}

This experiment showed that sound interactive analysis can prove the absence of data races in a complicated real-world application.
For ideal results, the heap-based analyses of the \textsc{Goblint} analyzer need to be improved in order to establish the one-to-one correspondence between
threads and dynamically allocated structs.
Nevertheless, the analyzer could be used to certify race-freedom for a fine-grained locking scheme in the program's default configuration
that increases parallelism.
Most importantly for our evaluation, we found that the reduction in time from almost 5 minutes to around 20 seconds makes it
much easier to attempt different approaches in order to certify race freedom.

\textsc{smtprc} (SMTP Open Relay Checker)\footnote{\url{https://sourceforge.net/projects/smtprc/}} is a program that contains real races. We analyze the latest released version
with \textsc{Goblint} configured similarly to the above, except we enable restarting of flow-insensitively unknowns (\cref{sec:restart}).
\begin{enumerate}
    \item The initial analysis takes 11 seconds and finds 30 races. Opening files in the GUI makes it easy to address these one-by-one. Opening the \verb|scan_engine| one finds a clear race on the structure that tracks the number of threads. The developer filters the warnings by \verb|o.current|.
    One finds that busy-looping code does reacquire the mutex after waking up from sleep. It is necessary to add the missing locking operation at the end of the loop.
    \item Upon saving, the analyzer can confirm that there is no longer a race on the loop's entry condition. The developer proceeds to protect racy accesses. Each such re-analysis takes 4 seconds in simple incremental mode; with restarting enabled, reanalysis takes between 6-8 seconds.
    \item The final access to this variable is a debug statement that is guarded by two boolean flags: debug and verbosity. As races in the logging can be considered benign, the developer experimentally overrides these flags to be false in order to confirm that the race is limited to these modes alone.
    \item Reanalysis without restart takes 4 seconds, but does not reveal a difference. With restarting, the reanalysis takes 6 seconds and verifies that the access is now dead. As this was the last unprotected access, the analyzer now reports 29 races.
\end{enumerate}
A screen recording of this workflow with the IDE integration is available in the artifact~\cite{artifact}.

We find that the incremental approach provides noticeable time savings and allows fixing one bug after the other.
As a data race consists of multiple accesses, the interactive experience allows the developer
to add synchronization one location at a time and near-instantly see error markers disappear for that access.
There still remained a high number of false alarms for this program,
which highlights the need to improve the analysis of non-locking synchronization patterns.
Evaluating the impact of interactivity, we found that even the more modest time savings when running with restarting
improved the experience when working within our IDE where reanalysis is triggered upon each save.

%% file: figures/precision_figure.pgf
\begingroup%
\makeatletter%
\begin{pgfpicture}%
\pgfpathrectangle{\pgfpointorigin}{\pgfqpoint{3.300000in}{2.200000in}}%
\pgfusepath{use as bounding box, clip}%
\begin{pgfscope}%
\pgfsetbuttcap%
\pgfsetmiterjoin%
\definecolor{currentfill}{rgb}{1.000000,1.000000,1.000000}%
\pgfsetfillcolor{currentfill}%
\pgfsetlinewidth{0.000000pt}%
\definecolor{currentstroke}{rgb}{1.000000,1.000000,1.000000}%
\pgfsetstrokecolor{currentstroke}%
\pgfsetdash{}{0pt}%
\pgfpathmoveto{\pgfqpoint{0.000000in}{0.000000in}}%
\pgfpathlineto{\pgfqpoint{3.300000in}{0.000000in}}%
\pgfpathlineto{\pgfqpoint{3.300000in}{2.200000in}}%
\pgfpathlineto{\pgfqpoint{0.000000in}{2.200000in}}%
\pgfpathlineto{\pgfqpoint{0.000000in}{0.000000in}}%
\pgfpathclose%
\pgfusepath{fill}%
\end{pgfscope}%
\begin{pgfscope}%
\pgfsetbuttcap%
\pgfsetmiterjoin%
\definecolor{currentfill}{rgb}{1.000000,1.000000,1.000000}%
\pgfsetfillcolor{currentfill}%
\pgfsetlinewidth{0.000000pt}%
\definecolor{currentstroke}{rgb}{0.000000,0.000000,0.000000}%
\pgfsetstrokecolor{currentstroke}%
\pgfsetstrokeopacity{0.000000}%
\pgfsetdash{}{0pt}%
\pgfpathmoveto{\pgfqpoint{0.542283in}{0.425000in}}%
\pgfpathlineto{\pgfqpoint{3.250000in}{0.425000in}}%
\pgfpathlineto{\pgfqpoint{3.250000in}{2.150000in}}%
\pgfpathlineto{\pgfqpoint{0.542283in}{2.150000in}}%
\pgfpathlineto{\pgfqpoint{0.542283in}{0.425000in}}%
\pgfpathclose%
\pgfusepath{fill}%
\end{pgfscope}%
\begin{pgfscope}%
\pgfsetbuttcap%
\pgfsetroundjoin%
\definecolor{currentfill}{rgb}{0.000000,0.000000,0.000000}%
\pgfsetfillcolor{currentfill}%
\pgfsetlinewidth{0.803000pt}%
\definecolor{currentstroke}{rgb}{0.000000,0.000000,0.000000}%
\pgfsetstrokecolor{currentstroke}%
\pgfsetdash{}{0pt}%
\pgfsys@defobject{currentmarker}{\pgfqpoint{0.000000in}{-0.048611in}}{\pgfqpoint{0.000000in}{0.000000in}}{%
\pgfpathmoveto{\pgfqpoint{0.000000in}{0.000000in}}%
\pgfpathlineto{\pgfqpoint{0.000000in}{-0.048611in}}%
\pgfusepath{stroke,fill}%
}%
\begin{pgfscope}%
\pgfsys@transformshift{1.212374in}{0.425000in}%
\pgfsys@useobject{currentmarker}{}%
\end{pgfscope}%
\end{pgfscope}%
\begin{pgfscope}%
\definecolor{textcolor}{rgb}{0.000000,0.000000,0.000000}%
\pgfsetstrokecolor{textcolor}%
\pgfsetfillcolor{textcolor}%
\pgftext[x=1.212374in,y=0.327778in,,top]{\color{textcolor}\rmfamily\fontsize{9.000000}{10.800000}\selectfont \(\displaystyle {5}\)}%
\end{pgfscope}%
\begin{pgfscope}%
\pgfsetbuttcap%
\pgfsetroundjoin%
\definecolor{currentfill}{rgb}{0.000000,0.000000,0.000000}%
\pgfsetfillcolor{currentfill}%
\pgfsetlinewidth{0.803000pt}%
\definecolor{currentstroke}{rgb}{0.000000,0.000000,0.000000}%
\pgfsetstrokecolor{currentstroke}%
\pgfsetdash{}{0pt}%
\pgfsys@defobject{currentmarker}{\pgfqpoint{0.000000in}{-0.048611in}}{\pgfqpoint{0.000000in}{0.000000in}}{%
\pgfpathmoveto{\pgfqpoint{0.000000in}{0.000000in}}%
\pgfpathlineto{\pgfqpoint{0.000000in}{-0.048611in}}%
\pgfusepath{stroke,fill}%
}%
\begin{pgfscope}%
\pgfsys@transformshift{1.896141in}{0.425000in}%
\pgfsys@useobject{currentmarker}{}%
\end{pgfscope}%
\end{pgfscope}%
\begin{pgfscope}%
\definecolor{textcolor}{rgb}{0.000000,0.000000,0.000000}%
\pgfsetstrokecolor{textcolor}%
\pgfsetfillcolor{textcolor}%
\pgftext[x=1.896141in,y=0.327778in,,top]{\color{textcolor}\rmfamily\fontsize{9.000000}{10.800000}\selectfont \(\displaystyle {10}\)}%
\end{pgfscope}%
\begin{pgfscope}%
\pgfsetbuttcap%
\pgfsetroundjoin%
\definecolor{currentfill}{rgb}{0.000000,0.000000,0.000000}%
\pgfsetfillcolor{currentfill}%
\pgfsetlinewidth{0.803000pt}%
\definecolor{currentstroke}{rgb}{0.000000,0.000000,0.000000}%
\pgfsetstrokecolor{currentstroke}%
\pgfsetdash{}{0pt}%
\pgfsys@defobject{currentmarker}{\pgfqpoint{0.000000in}{-0.048611in}}{\pgfqpoint{0.000000in}{0.000000in}}{%
\pgfpathmoveto{\pgfqpoint{0.000000in}{0.000000in}}%
\pgfpathlineto{\pgfqpoint{0.000000in}{-0.048611in}}%
\pgfusepath{stroke,fill}%
}%
\begin{pgfscope}%
\pgfsys@transformshift{2.579908in}{0.425000in}%
\pgfsys@useobject{currentmarker}{}%
\end{pgfscope}%
\end{pgfscope}%
\begin{pgfscope}%
\definecolor{textcolor}{rgb}{0.000000,0.000000,0.000000}%
\pgfsetstrokecolor{textcolor}%
\pgfsetfillcolor{textcolor}%
\pgftext[x=2.579908in,y=0.327778in,,top]{\color{textcolor}\rmfamily\fontsize{9.000000}{10.800000}\selectfont \(\displaystyle {15}\)}%
\end{pgfscope}%
\begin{pgfscope}%
\definecolor{textcolor}{rgb}{0.000000,0.000000,0.000000}%
\pgfsetstrokecolor{textcolor}%
\pgfsetfillcolor{textcolor}%
\pgftext[x=1.896141in,y=0.161111in,,top]{\color{textcolor}\rmfamily\fontsize{9.000000}{10.800000}\selectfont \# Commits}%
\end{pgfscope}%
\begin{pgfscope}%
\pgfsetbuttcap%
\pgfsetroundjoin%
\definecolor{currentfill}{rgb}{0.000000,0.000000,0.000000}%
\pgfsetfillcolor{currentfill}%
\pgfsetlinewidth{0.803000pt}%
\definecolor{currentstroke}{rgb}{0.000000,0.000000,0.000000}%
\pgfsetstrokecolor{currentstroke}%
\pgfsetdash{}{0pt}%
\pgfsys@defobject{currentmarker}{\pgfqpoint{-0.048611in}{0.000000in}}{\pgfqpoint{-0.000000in}{0.000000in}}{%
\pgfpathmoveto{\pgfqpoint{-0.000000in}{0.000000in}}%
\pgfpathlineto{\pgfqpoint{-0.048611in}{0.000000in}}%
\pgfusepath{stroke,fill}%
}%
\begin{pgfscope}%
\pgfsys@transformshift{0.542283in}{0.469231in}%
\pgfsys@useobject{currentmarker}{}%
\end{pgfscope}%
\end{pgfscope}%
\begin{pgfscope}%
\definecolor{textcolor}{rgb}{0.000000,0.000000,0.000000}%
\pgfsetstrokecolor{textcolor}%
\pgfsetfillcolor{textcolor}%
\pgftext[x=0.216667in, y=0.425828in, left, base]{\color{textcolor}\rmfamily\fontsize{9.000000}{10.800000}\selectfont \(\displaystyle {0.00}\)}%
\end{pgfscope}%
\begin{pgfscope}%
\pgfsetbuttcap%
\pgfsetroundjoin%
\definecolor{currentfill}{rgb}{0.000000,0.000000,0.000000}%
\pgfsetfillcolor{currentfill}%
\pgfsetlinewidth{0.803000pt}%
\definecolor{currentstroke}{rgb}{0.000000,0.000000,0.000000}%
\pgfsetstrokecolor{currentstroke}%
\pgfsetdash{}{0pt}%
\pgfsys@defobject{currentmarker}{\pgfqpoint{-0.048611in}{0.000000in}}{\pgfqpoint{-0.000000in}{0.000000in}}{%
\pgfpathmoveto{\pgfqpoint{-0.000000in}{0.000000in}}%
\pgfpathlineto{\pgfqpoint{-0.048611in}{0.000000in}}%
\pgfusepath{stroke,fill}%
}%
\begin{pgfscope}%
\pgfsys@transformshift{0.542283in}{0.911538in}%
\pgfsys@useobject{currentmarker}{}%
\end{pgfscope}%
\end{pgfscope}%
\begin{pgfscope}%
\definecolor{textcolor}{rgb}{0.000000,0.000000,0.000000}%
\pgfsetstrokecolor{textcolor}%
\pgfsetfillcolor{textcolor}%
\pgftext[x=0.216667in, y=0.868136in, left, base]{\color{textcolor}\rmfamily\fontsize{9.000000}{10.800000}\selectfont \(\displaystyle {0.05}\)}%
\end{pgfscope}%
\begin{pgfscope}%
\pgfsetbuttcap%
\pgfsetroundjoin%
\definecolor{currentfill}{rgb}{0.000000,0.000000,0.000000}%
\pgfsetfillcolor{currentfill}%
\pgfsetlinewidth{0.803000pt}%
\definecolor{currentstroke}{rgb}{0.000000,0.000000,0.000000}%
\pgfsetstrokecolor{currentstroke}%
\pgfsetdash{}{0pt}%
\pgfsys@defobject{currentmarker}{\pgfqpoint{-0.048611in}{0.000000in}}{\pgfqpoint{-0.000000in}{0.000000in}}{%
\pgfpathmoveto{\pgfqpoint{-0.000000in}{0.000000in}}%
\pgfpathlineto{\pgfqpoint{-0.048611in}{0.000000in}}%
\pgfusepath{stroke,fill}%
}%
\begin{pgfscope}%
\pgfsys@transformshift{0.542283in}{1.353846in}%
\pgfsys@useobject{currentmarker}{}%
\end{pgfscope}%
\end{pgfscope}%
\begin{pgfscope}%
\definecolor{textcolor}{rgb}{0.000000,0.000000,0.000000}%
\pgfsetstrokecolor{textcolor}%
\pgfsetfillcolor{textcolor}%
\pgftext[x=0.216667in, y=1.310443in, left, base]{\color{textcolor}\rmfamily\fontsize{9.000000}{10.800000}\selectfont \(\displaystyle {0.10}\)}%
\end{pgfscope}%
\begin{pgfscope}%
\pgfsetbuttcap%
\pgfsetroundjoin%
\definecolor{currentfill}{rgb}{0.000000,0.000000,0.000000}%
\pgfsetfillcolor{currentfill}%
\pgfsetlinewidth{0.803000pt}%
\definecolor{currentstroke}{rgb}{0.000000,0.000000,0.000000}%
\pgfsetstrokecolor{currentstroke}%
\pgfsetdash{}{0pt}%
\pgfsys@defobject{currentmarker}{\pgfqpoint{-0.048611in}{0.000000in}}{\pgfqpoint{-0.000000in}{0.000000in}}{%
\pgfpathmoveto{\pgfqpoint{-0.000000in}{0.000000in}}%
\pgfpathlineto{\pgfqpoint{-0.048611in}{0.000000in}}%
\pgfusepath{stroke,fill}%
}%
\begin{pgfscope}%
\pgfsys@transformshift{0.542283in}{1.796154in}%
\pgfsys@useobject{currentmarker}{}%
\end{pgfscope}%
\end{pgfscope}%
\begin{pgfscope}%
\definecolor{textcolor}{rgb}{0.000000,0.000000,0.000000}%
\pgfsetstrokecolor{textcolor}%
\pgfsetfillcolor{textcolor}%
\pgftext[x=0.216667in, y=1.752751in, left, base]{\color{textcolor}\rmfamily\fontsize{9.000000}{10.800000}\selectfont \(\displaystyle {0.15}\)}%
\end{pgfscope}%
\begin{pgfscope}%
\definecolor{textcolor}{rgb}{0.000000,0.000000,0.000000}%
\pgfsetstrokecolor{textcolor}%
\pgfsetfillcolor{textcolor}%
\pgftext[x=0.161111in,y=1.287500in,,bottom,rotate=90.000000]{\color{textcolor}\rmfamily\fontsize{9.000000}{10.800000}\scriptsize Share of less precise program points}%
\end{pgfscope}%
\begin{pgfscope}%
\pgfpathrectangle{\pgfqpoint{0.542283in}{0.425000in}}{\pgfqpoint{2.707717in}{1.725000in}}%
\pgfusepath{clip}%
\pgfsetbuttcap%
\pgfsetroundjoin%
\pgfsetlinewidth{0.401500pt}%
\definecolor{currentstroke}{rgb}{1.000000,0.000000,0.000000}%
\pgfsetstrokecolor{currentstroke}%
\pgfsetstrokeopacity{0.850000}%
\pgfsetdash{{1.480000pt}{0.640000pt}}{0.000000pt}%
\pgfpathmoveto{\pgfqpoint{0.665361in}{0.469231in}}%
\pgfpathlineto{\pgfqpoint{0.802114in}{0.660563in}}%
\pgfpathlineto{\pgfqpoint{1.212374in}{0.539259in}}%
\pgfpathlineto{\pgfqpoint{1.485881in}{0.539259in}}%
\pgfusepath{stroke}%
\end{pgfscope}%
\begin{pgfscope}%
\pgfpathrectangle{\pgfqpoint{0.542283in}{0.425000in}}{\pgfqpoint{2.707717in}{1.725000in}}%
\pgfusepath{clip}%
\pgfsetbuttcap%
\pgfsetroundjoin%
\definecolor{currentfill}{rgb}{1.000000,0.000000,0.000000}%
\pgfsetfillcolor{currentfill}%
\pgfsetfillopacity{0.850000}%
\pgfsetlinewidth{1.003750pt}%
\definecolor{currentstroke}{rgb}{1.000000,0.000000,0.000000}%
\pgfsetstrokecolor{currentstroke}%
\pgfsetstrokeopacity{0.850000}%
\pgfsetdash{}{0pt}%
\pgfsys@defobject{currentmarker}{\pgfqpoint{-0.006944in}{-0.006944in}}{\pgfqpoint{0.006944in}{0.006944in}}{%
\pgfpathmoveto{\pgfqpoint{-0.006944in}{-0.006944in}}%
\pgfpathlineto{\pgfqpoint{0.006944in}{0.006944in}}%
\pgfpathmoveto{\pgfqpoint{-0.006944in}{0.006944in}}%
\pgfpathlineto{\pgfqpoint{0.006944in}{-0.006944in}}%
\pgfusepath{stroke,fill}%
}%
\begin{pgfscope}%
\pgfsys@transformshift{0.665361in}{0.469231in}%
\pgfsys@useobject{currentmarker}{}%
\end{pgfscope}%
\begin{pgfscope}%
\pgfsys@transformshift{0.802114in}{0.660563in}%
\pgfsys@useobject{currentmarker}{}%
\end{pgfscope}%
\begin{pgfscope}%
\pgfsys@transformshift{1.212374in}{0.539259in}%
\pgfsys@useobject{currentmarker}{}%
\end{pgfscope}%
\begin{pgfscope}%
\pgfsys@transformshift{1.485881in}{0.539259in}%
\pgfsys@useobject{currentmarker}{}%
\end{pgfscope}%
\end{pgfscope}%
\begin{pgfscope}%
\pgfpathrectangle{\pgfqpoint{0.542283in}{0.425000in}}{\pgfqpoint{2.707717in}{1.725000in}}%
\pgfusepath{clip}%
\pgfsetbuttcap%
\pgfsetroundjoin%
\pgfsetlinewidth{0.401500pt}%
\definecolor{currentstroke}{rgb}{0.941176,1.000000,1.000000}%
\pgfsetstrokecolor{currentstroke}%
\pgfsetstrokeopacity{0.850000}%
\pgfsetdash{{1.480000pt}{0.640000pt}}{0.000000pt}%
\pgfpathmoveto{\pgfqpoint{0.665361in}{0.469231in}}%
\pgfpathlineto{\pgfqpoint{0.802114in}{0.469231in}}%
\pgfpathlineto{\pgfqpoint{1.212374in}{0.469231in}}%
\pgfpathlineto{\pgfqpoint{1.485881in}{2.004024in}}%
\pgfusepath{stroke}%
\end{pgfscope}%
\begin{pgfscope}%
\pgfpathrectangle{\pgfqpoint{0.542283in}{0.425000in}}{\pgfqpoint{2.707717in}{1.725000in}}%
\pgfusepath{clip}%
\pgfsetbuttcap%
\pgfsetroundjoin%
\definecolor{currentfill}{rgb}{0.941176,1.000000,1.000000}%
\pgfsetfillcolor{currentfill}%
\pgfsetfillopacity{0.850000}%
\pgfsetlinewidth{1.003750pt}%
\definecolor{currentstroke}{rgb}{0.941176,1.000000,1.000000}%
\pgfsetstrokecolor{currentstroke}%
\pgfsetstrokeopacity{0.850000}%
\pgfsetdash{}{0pt}%
\pgfsys@defobject{currentmarker}{\pgfqpoint{-0.006944in}{-0.006944in}}{\pgfqpoint{0.006944in}{0.006944in}}{%
\pgfpathmoveto{\pgfqpoint{-0.006944in}{-0.006944in}}%
\pgfpathlineto{\pgfqpoint{0.006944in}{0.006944in}}%
\pgfpathmoveto{\pgfqpoint{-0.006944in}{0.006944in}}%
\pgfpathlineto{\pgfqpoint{0.006944in}{-0.006944in}}%
\pgfusepath{stroke,fill}%
}%
\begin{pgfscope}%
\pgfsys@transformshift{0.665361in}{0.469231in}%
\pgfsys@useobject{currentmarker}{}%
\end{pgfscope}%
\begin{pgfscope}%
\pgfsys@transformshift{0.802114in}{0.469231in}%
\pgfsys@useobject{currentmarker}{}%
\end{pgfscope}%
\begin{pgfscope}%
\pgfsys@transformshift{1.212374in}{0.469231in}%
\pgfsys@useobject{currentmarker}{}%
\end{pgfscope}%
\begin{pgfscope}%
\pgfsys@transformshift{1.485881in}{2.004024in}%
\pgfsys@useobject{currentmarker}{}%
\end{pgfscope}%
\end{pgfscope}%
\begin{pgfscope}%
\pgfpathrectangle{\pgfqpoint{0.542283in}{0.425000in}}{\pgfqpoint{2.707717in}{1.725000in}}%
\pgfusepath{clip}%
\pgfsetbuttcap%
\pgfsetroundjoin%
\pgfsetlinewidth{0.401500pt}%
\definecolor{currentstroke}{rgb}{0.000000,0.000000,1.000000}%
\pgfsetstrokecolor{currentstroke}%
\pgfsetstrokeopacity{0.850000}%
\pgfsetdash{{1.480000pt}{0.640000pt}}{0.000000pt}%
\pgfpathmoveto{\pgfqpoint{0.665361in}{0.530240in}}%
\pgfpathlineto{\pgfqpoint{0.802114in}{0.533502in}}%
\pgfpathlineto{\pgfqpoint{1.896141in}{0.812058in}}%
\pgfusepath{stroke}%
\end{pgfscope}%
\begin{pgfscope}%
\pgfpathrectangle{\pgfqpoint{0.542283in}{0.425000in}}{\pgfqpoint{2.707717in}{1.725000in}}%
\pgfusepath{clip}%
\pgfsetbuttcap%
\pgfsetroundjoin%
\definecolor{currentfill}{rgb}{0.000000,0.000000,1.000000}%
\pgfsetfillcolor{currentfill}%
\pgfsetfillopacity{0.850000}%
\pgfsetlinewidth{1.003750pt}%
\definecolor{currentstroke}{rgb}{0.000000,0.000000,1.000000}%
\pgfsetstrokecolor{currentstroke}%
\pgfsetstrokeopacity{0.850000}%
\pgfsetdash{}{0pt}%
\pgfsys@defobject{currentmarker}{\pgfqpoint{-0.006944in}{-0.006944in}}{\pgfqpoint{0.006944in}{0.006944in}}{%
\pgfpathmoveto{\pgfqpoint{-0.006944in}{-0.006944in}}%
\pgfpathlineto{\pgfqpoint{0.006944in}{0.006944in}}%
\pgfpathmoveto{\pgfqpoint{-0.006944in}{0.006944in}}%
\pgfpathlineto{\pgfqpoint{0.006944in}{-0.006944in}}%
\pgfusepath{stroke,fill}%
}%
\begin{pgfscope}%
\pgfsys@transformshift{0.665361in}{0.530240in}%
\pgfsys@useobject{currentmarker}{}%
\end{pgfscope}%
\begin{pgfscope}%
\pgfsys@transformshift{0.802114in}{0.533502in}%
\pgfsys@useobject{currentmarker}{}%
\end{pgfscope}%
\begin{pgfscope}%
\pgfsys@transformshift{1.896141in}{0.812058in}%
\pgfsys@useobject{currentmarker}{}%
\end{pgfscope}%
\end{pgfscope}%
\begin{pgfscope}%
\pgfpathrectangle{\pgfqpoint{0.542283in}{0.425000in}}{\pgfqpoint{2.707717in}{1.725000in}}%
\pgfusepath{clip}%
\pgfsetbuttcap%
\pgfsetroundjoin%
\pgfsetlinewidth{0.401500pt}%
\definecolor{currentstroke}{rgb}{0.647059,0.164706,0.164706}%
\pgfsetstrokecolor{currentstroke}%
\pgfsetstrokeopacity{0.850000}%
\pgfsetdash{{1.480000pt}{0.640000pt}}{0.000000pt}%
\pgfpathmoveto{\pgfqpoint{0.665361in}{0.469231in}}%
\pgfpathlineto{\pgfqpoint{0.802114in}{0.477533in}}%
\pgfpathlineto{\pgfqpoint{1.212374in}{0.487717in}}%
\pgfpathlineto{\pgfqpoint{1.622634in}{0.487717in}}%
\pgfusepath{stroke}%
\end{pgfscope}%
\begin{pgfscope}%
\pgfpathrectangle{\pgfqpoint{0.542283in}{0.425000in}}{\pgfqpoint{2.707717in}{1.725000in}}%
\pgfusepath{clip}%
\pgfsetbuttcap%
\pgfsetroundjoin%
\definecolor{currentfill}{rgb}{0.647059,0.164706,0.164706}%
\pgfsetfillcolor{currentfill}%
\pgfsetfillopacity{0.850000}%
\pgfsetlinewidth{1.003750pt}%
\definecolor{currentstroke}{rgb}{0.647059,0.164706,0.164706}%
\pgfsetstrokecolor{currentstroke}%
\pgfsetstrokeopacity{0.850000}%
\pgfsetdash{}{0pt}%
\pgfsys@defobject{currentmarker}{\pgfqpoint{-0.006944in}{-0.006944in}}{\pgfqpoint{0.006944in}{0.006944in}}{%
\pgfpathmoveto{\pgfqpoint{-0.006944in}{-0.006944in}}%
\pgfpathlineto{\pgfqpoint{0.006944in}{0.006944in}}%
\pgfpathmoveto{\pgfqpoint{-0.006944in}{0.006944in}}%
\pgfpathlineto{\pgfqpoint{0.006944in}{-0.006944in}}%
\pgfusepath{stroke,fill}%
}%
\begin{pgfscope}%
\pgfsys@transformshift{0.665361in}{0.469231in}%
\pgfsys@useobject{currentmarker}{}%
\end{pgfscope}%
\begin{pgfscope}%
\pgfsys@transformshift{0.802114in}{0.477533in}%
\pgfsys@useobject{currentmarker}{}%
\end{pgfscope}%
\begin{pgfscope}%
\pgfsys@transformshift{1.212374in}{0.487717in}%
\pgfsys@useobject{currentmarker}{}%
\end{pgfscope}%
\begin{pgfscope}%
\pgfsys@transformshift{1.622634in}{0.487717in}%
\pgfsys@useobject{currentmarker}{}%
\end{pgfscope}%
\end{pgfscope}%
\begin{pgfscope}%
\pgfpathrectangle{\pgfqpoint{0.542283in}{0.425000in}}{\pgfqpoint{2.707717in}{1.725000in}}%
\pgfusepath{clip}%
\pgfsetbuttcap%
\pgfsetroundjoin%
\pgfsetlinewidth{0.401500pt}%
\definecolor{currentstroke}{rgb}{0.498039,1.000000,0.000000}%
\pgfsetstrokecolor{currentstroke}%
\pgfsetstrokeopacity{0.850000}%
\pgfsetdash{{1.480000pt}{0.640000pt}}{0.000000pt}%
\pgfpathmoveto{\pgfqpoint{0.665361in}{0.469231in}}%
\pgfpathlineto{\pgfqpoint{0.802114in}{0.469231in}}%
\pgfpathlineto{\pgfqpoint{1.212374in}{0.469231in}}%
\pgfpathlineto{\pgfqpoint{1.349128in}{0.469231in}}%
\pgfusepath{stroke}%
\end{pgfscope}%
\begin{pgfscope}%
\pgfpathrectangle{\pgfqpoint{0.542283in}{0.425000in}}{\pgfqpoint{2.707717in}{1.725000in}}%
\pgfusepath{clip}%
\pgfsetbuttcap%
\pgfsetroundjoin%
\definecolor{currentfill}{rgb}{0.498039,1.000000,0.000000}%
\pgfsetfillcolor{currentfill}%
\pgfsetfillopacity{0.850000}%
\pgfsetlinewidth{1.003750pt}%
\definecolor{currentstroke}{rgb}{0.498039,1.000000,0.000000}%
\pgfsetstrokecolor{currentstroke}%
\pgfsetstrokeopacity{0.850000}%
\pgfsetdash{}{0pt}%
\pgfsys@defobject{currentmarker}{\pgfqpoint{-0.006944in}{-0.006944in}}{\pgfqpoint{0.006944in}{0.006944in}}{%
\pgfpathmoveto{\pgfqpoint{-0.006944in}{-0.006944in}}%
\pgfpathlineto{\pgfqpoint{0.006944in}{0.006944in}}%
\pgfpathmoveto{\pgfqpoint{-0.006944in}{0.006944in}}%
\pgfpathlineto{\pgfqpoint{0.006944in}{-0.006944in}}%
\pgfusepath{stroke,fill}%
}%
\begin{pgfscope}%
\pgfsys@transformshift{0.665361in}{0.469231in}%
\pgfsys@useobject{currentmarker}{}%
\end{pgfscope}%
\begin{pgfscope}%
\pgfsys@transformshift{0.802114in}{0.469231in}%
\pgfsys@useobject{currentmarker}{}%
\end{pgfscope}%
\begin{pgfscope}%
\pgfsys@transformshift{1.212374in}{0.469231in}%
\pgfsys@useobject{currentmarker}{}%
\end{pgfscope}%
\begin{pgfscope}%
\pgfsys@transformshift{1.349128in}{0.469231in}%
\pgfsys@useobject{currentmarker}{}%
\end{pgfscope}%
\end{pgfscope}%
\begin{pgfscope}%
\pgfpathrectangle{\pgfqpoint{0.542283in}{0.425000in}}{\pgfqpoint{2.707717in}{1.725000in}}%
\pgfusepath{clip}%
\pgfsetbuttcap%
\pgfsetroundjoin%
\pgfsetlinewidth{0.401500pt}%
\definecolor{currentstroke}{rgb}{0.823529,0.411765,0.117647}%
\pgfsetstrokecolor{currentstroke}%
\pgfsetstrokeopacity{0.850000}%
\pgfsetdash{{1.480000pt}{0.640000pt}}{0.000000pt}%
\pgfpathmoveto{\pgfqpoint{0.665361in}{1.062791in}}%
\pgfpathlineto{\pgfqpoint{0.802114in}{1.078412in}}%
\pgfpathlineto{\pgfqpoint{1.212374in}{1.071373in}}%
\pgfpathlineto{\pgfqpoint{1.896141in}{1.071373in}}%
\pgfpathlineto{\pgfqpoint{2.579908in}{1.071105in}}%
\pgfpathlineto{\pgfqpoint{2.853415in}{1.071105in}}%
\pgfusepath{stroke}%
\end{pgfscope}%
\begin{pgfscope}%
\pgfpathrectangle{\pgfqpoint{0.542283in}{0.425000in}}{\pgfqpoint{2.707717in}{1.725000in}}%
\pgfusepath{clip}%
\pgfsetbuttcap%
\pgfsetroundjoin%
\definecolor{currentfill}{rgb}{0.823529,0.411765,0.117647}%
\pgfsetfillcolor{currentfill}%
\pgfsetfillopacity{0.850000}%
\pgfsetlinewidth{1.003750pt}%
\definecolor{currentstroke}{rgb}{0.823529,0.411765,0.117647}%
\pgfsetstrokecolor{currentstroke}%
\pgfsetstrokeopacity{0.850000}%
\pgfsetdash{}{0pt}%
\pgfsys@defobject{currentmarker}{\pgfqpoint{-0.006944in}{-0.006944in}}{\pgfqpoint{0.006944in}{0.006944in}}{%
\pgfpathmoveto{\pgfqpoint{-0.006944in}{-0.006944in}}%
\pgfpathlineto{\pgfqpoint{0.006944in}{0.006944in}}%
\pgfpathmoveto{\pgfqpoint{-0.006944in}{0.006944in}}%
\pgfpathlineto{\pgfqpoint{0.006944in}{-0.006944in}}%
\pgfusepath{stroke,fill}%
}%
\begin{pgfscope}%
\pgfsys@transformshift{0.665361in}{1.062791in}%
\pgfsys@useobject{currentmarker}{}%
\end{pgfscope}%
\begin{pgfscope}%
\pgfsys@transformshift{0.802114in}{1.078412in}%
\pgfsys@useobject{currentmarker}{}%
\end{pgfscope}%
\begin{pgfscope}%
\pgfsys@transformshift{1.212374in}{1.071373in}%
\pgfsys@useobject{currentmarker}{}%
\end{pgfscope}%
\begin{pgfscope}%
\pgfsys@transformshift{1.896141in}{1.071373in}%
\pgfsys@useobject{currentmarker}{}%
\end{pgfscope}%
\begin{pgfscope}%
\pgfsys@transformshift{2.579908in}{1.071105in}%
\pgfsys@useobject{currentmarker}{}%
\end{pgfscope}%
\begin{pgfscope}%
\pgfsys@transformshift{2.853415in}{1.071105in}%
\pgfsys@useobject{currentmarker}{}%
\end{pgfscope}%
\end{pgfscope}%
\begin{pgfscope}%
\pgfpathrectangle{\pgfqpoint{0.542283in}{0.425000in}}{\pgfqpoint{2.707717in}{1.725000in}}%
\pgfusepath{clip}%
\pgfsetbuttcap%
\pgfsetroundjoin%
\pgfsetlinewidth{0.401500pt}%
\definecolor{currentstroke}{rgb}{0.000000,0.000000,0.545098}%
\pgfsetstrokecolor{currentstroke}%
\pgfsetstrokeopacity{0.850000}%
\pgfsetdash{{1.480000pt}{0.640000pt}}{0.000000pt}%
\pgfpathmoveto{\pgfqpoint{0.665361in}{0.675561in}}%
\pgfpathlineto{\pgfqpoint{0.802114in}{0.570146in}}%
\pgfpathlineto{\pgfqpoint{1.212374in}{0.571564in}}%
\pgfpathlineto{\pgfqpoint{1.896141in}{0.570137in}}%
\pgfpathlineto{\pgfqpoint{2.443155in}{0.574164in}}%
\pgfusepath{stroke}%
\end{pgfscope}%
\begin{pgfscope}%
\pgfpathrectangle{\pgfqpoint{0.542283in}{0.425000in}}{\pgfqpoint{2.707717in}{1.725000in}}%
\pgfusepath{clip}%
\pgfsetbuttcap%
\pgfsetroundjoin%
\definecolor{currentfill}{rgb}{0.000000,0.000000,0.545098}%
\pgfsetfillcolor{currentfill}%
\pgfsetfillopacity{0.850000}%
\pgfsetlinewidth{1.003750pt}%
\definecolor{currentstroke}{rgb}{0.000000,0.000000,0.545098}%
\pgfsetstrokecolor{currentstroke}%
\pgfsetstrokeopacity{0.850000}%
\pgfsetdash{}{0pt}%
\pgfsys@defobject{currentmarker}{\pgfqpoint{-0.006944in}{-0.006944in}}{\pgfqpoint{0.006944in}{0.006944in}}{%
\pgfpathmoveto{\pgfqpoint{-0.006944in}{-0.006944in}}%
\pgfpathlineto{\pgfqpoint{0.006944in}{0.006944in}}%
\pgfpathmoveto{\pgfqpoint{-0.006944in}{0.006944in}}%
\pgfpathlineto{\pgfqpoint{0.006944in}{-0.006944in}}%
\pgfusepath{stroke,fill}%
}%
\begin{pgfscope}%
\pgfsys@transformshift{0.665361in}{0.675561in}%
\pgfsys@useobject{currentmarker}{}%
\end{pgfscope}%
\begin{pgfscope}%
\pgfsys@transformshift{0.802114in}{0.570146in}%
\pgfsys@useobject{currentmarker}{}%
\end{pgfscope}%
\begin{pgfscope}%
\pgfsys@transformshift{1.212374in}{0.571564in}%
\pgfsys@useobject{currentmarker}{}%
\end{pgfscope}%
\begin{pgfscope}%
\pgfsys@transformshift{1.896141in}{0.570137in}%
\pgfsys@useobject{currentmarker}{}%
\end{pgfscope}%
\begin{pgfscope}%
\pgfsys@transformshift{2.443155in}{0.574164in}%
\pgfsys@useobject{currentmarker}{}%
\end{pgfscope}%
\end{pgfscope}%
\begin{pgfscope}%
\pgfpathrectangle{\pgfqpoint{0.542283in}{0.425000in}}{\pgfqpoint{2.707717in}{1.725000in}}%
\pgfusepath{clip}%
\pgfsetbuttcap%
\pgfsetroundjoin%
\pgfsetlinewidth{0.401500pt}%
\definecolor{currentstroke}{rgb}{0.000000,0.392157,0.000000}%
\pgfsetstrokecolor{currentstroke}%
\pgfsetstrokeopacity{0.850000}%
\pgfsetdash{{1.480000pt}{0.640000pt}}{0.000000pt}%
\pgfpathmoveto{\pgfqpoint{0.665361in}{0.469231in}}%
\pgfpathlineto{\pgfqpoint{0.802114in}{0.469231in}}%
\pgfpathlineto{\pgfqpoint{1.212374in}{0.477830in}}%
\pgfpathlineto{\pgfqpoint{1.485881in}{0.477830in}}%
\pgfusepath{stroke}%
\end{pgfscope}%
\begin{pgfscope}%
\pgfpathrectangle{\pgfqpoint{0.542283in}{0.425000in}}{\pgfqpoint{2.707717in}{1.725000in}}%
\pgfusepath{clip}%
\pgfsetbuttcap%
\pgfsetroundjoin%
\definecolor{currentfill}{rgb}{0.000000,0.392157,0.000000}%
\pgfsetfillcolor{currentfill}%
\pgfsetfillopacity{0.850000}%
\pgfsetlinewidth{1.003750pt}%
\definecolor{currentstroke}{rgb}{0.000000,0.392157,0.000000}%
\pgfsetstrokecolor{currentstroke}%
\pgfsetstrokeopacity{0.850000}%
\pgfsetdash{}{0pt}%
\pgfsys@defobject{currentmarker}{\pgfqpoint{-0.006944in}{-0.006944in}}{\pgfqpoint{0.006944in}{0.006944in}}{%
\pgfpathmoveto{\pgfqpoint{-0.006944in}{-0.006944in}}%
\pgfpathlineto{\pgfqpoint{0.006944in}{0.006944in}}%
\pgfpathmoveto{\pgfqpoint{-0.006944in}{0.006944in}}%
\pgfpathlineto{\pgfqpoint{0.006944in}{-0.006944in}}%
\pgfusepath{stroke,fill}%
}%
\begin{pgfscope}%
\pgfsys@transformshift{0.665361in}{0.469231in}%
\pgfsys@useobject{currentmarker}{}%
\end{pgfscope}%
\begin{pgfscope}%
\pgfsys@transformshift{0.802114in}{0.469231in}%
\pgfsys@useobject{currentmarker}{}%
\end{pgfscope}%
\begin{pgfscope}%
\pgfsys@transformshift{1.212374in}{0.477830in}%
\pgfsys@useobject{currentmarker}{}%
\end{pgfscope}%
\begin{pgfscope}%
\pgfsys@transformshift{1.485881in}{0.477830in}%
\pgfsys@useobject{currentmarker}{}%
\end{pgfscope}%
\end{pgfscope}%
\begin{pgfscope}%
\pgfpathrectangle{\pgfqpoint{0.542283in}{0.425000in}}{\pgfqpoint{2.707717in}{1.725000in}}%
\pgfusepath{clip}%
\pgfsetbuttcap%
\pgfsetroundjoin%
\pgfsetlinewidth{0.401500pt}%
\definecolor{currentstroke}{rgb}{0.180392,0.545098,0.341176}%
\pgfsetstrokecolor{currentstroke}%
\pgfsetstrokeopacity{0.850000}%
\pgfsetdash{{1.480000pt}{0.640000pt}}{0.000000pt}%
\pgfpathmoveto{\pgfqpoint{0.665361in}{1.061366in}}%
\pgfpathlineto{\pgfqpoint{0.689570in}{2.153333in}}%
\pgfusepath{stroke}%
\end{pgfscope}%
\begin{pgfscope}%
\pgfpathrectangle{\pgfqpoint{0.542283in}{0.425000in}}{\pgfqpoint{2.707717in}{1.725000in}}%
\pgfusepath{clip}%
\pgfsetbuttcap%
\pgfsetroundjoin%
\definecolor{currentfill}{rgb}{0.180392,0.545098,0.341176}%
\pgfsetfillcolor{currentfill}%
\pgfsetfillopacity{0.850000}%
\pgfsetlinewidth{1.003750pt}%
\definecolor{currentstroke}{rgb}{0.180392,0.545098,0.341176}%
\pgfsetstrokecolor{currentstroke}%
\pgfsetstrokeopacity{0.850000}%
\pgfsetdash{}{0pt}%
\pgfsys@defobject{currentmarker}{\pgfqpoint{-0.006944in}{-0.006944in}}{\pgfqpoint{0.006944in}{0.006944in}}{%
\pgfpathmoveto{\pgfqpoint{-0.006944in}{-0.006944in}}%
\pgfpathlineto{\pgfqpoint{0.006944in}{0.006944in}}%
\pgfpathmoveto{\pgfqpoint{-0.006944in}{0.006944in}}%
\pgfpathlineto{\pgfqpoint{0.006944in}{-0.006944in}}%
\pgfusepath{stroke,fill}%
}%
\begin{pgfscope}%
\pgfsys@transformshift{0.665361in}{1.061366in}%
\pgfsys@useobject{currentmarker}{}%
\end{pgfscope}%
\begin{pgfscope}%
\pgfsys@transformshift{0.802114in}{7.229644in}%
\pgfsys@useobject{currentmarker}{}%
\end{pgfscope}%
\end{pgfscope}%
\begin{pgfscope}%
\pgfpathrectangle{\pgfqpoint{0.542283in}{0.425000in}}{\pgfqpoint{2.707717in}{1.725000in}}%
\pgfusepath{clip}%
\pgfsetbuttcap%
\pgfsetroundjoin%
\pgfsetlinewidth{0.401500pt}%
\definecolor{currentstroke}{rgb}{0.000000,0.501961,0.000000}%
\pgfsetstrokecolor{currentstroke}%
\pgfsetstrokeopacity{0.850000}%
\pgfsetdash{{1.480000pt}{0.640000pt}}{0.000000pt}%
\pgfpathmoveto{\pgfqpoint{0.665361in}{0.667975in}}%
\pgfpathlineto{\pgfqpoint{0.802114in}{0.678475in}}%
\pgfpathlineto{\pgfqpoint{1.212374in}{0.678236in}}%
\pgfpathlineto{\pgfqpoint{1.759388in}{0.678236in}}%
\pgfusepath{stroke}%
\end{pgfscope}%
\begin{pgfscope}%
\pgfpathrectangle{\pgfqpoint{0.542283in}{0.425000in}}{\pgfqpoint{2.707717in}{1.725000in}}%
\pgfusepath{clip}%
\pgfsetbuttcap%
\pgfsetroundjoin%
\definecolor{currentfill}{rgb}{0.000000,0.501961,0.000000}%
\pgfsetfillcolor{currentfill}%
\pgfsetfillopacity{0.850000}%
\pgfsetlinewidth{1.003750pt}%
\definecolor{currentstroke}{rgb}{0.000000,0.501961,0.000000}%
\pgfsetstrokecolor{currentstroke}%
\pgfsetstrokeopacity{0.850000}%
\pgfsetdash{}{0pt}%
\pgfsys@defobject{currentmarker}{\pgfqpoint{-0.006944in}{-0.006944in}}{\pgfqpoint{0.006944in}{0.006944in}}{%
\pgfpathmoveto{\pgfqpoint{-0.006944in}{-0.006944in}}%
\pgfpathlineto{\pgfqpoint{0.006944in}{0.006944in}}%
\pgfpathmoveto{\pgfqpoint{-0.006944in}{0.006944in}}%
\pgfpathlineto{\pgfqpoint{0.006944in}{-0.006944in}}%
\pgfusepath{stroke,fill}%
}%
\begin{pgfscope}%
\pgfsys@transformshift{0.665361in}{0.667975in}%
\pgfsys@useobject{currentmarker}{}%
\end{pgfscope}%
\begin{pgfscope}%
\pgfsys@transformshift{0.802114in}{0.678475in}%
\pgfsys@useobject{currentmarker}{}%
\end{pgfscope}%
\begin{pgfscope}%
\pgfsys@transformshift{1.212374in}{0.678236in}%
\pgfsys@useobject{currentmarker}{}%
\end{pgfscope}%
\begin{pgfscope}%
\pgfsys@transformshift{1.759388in}{0.678236in}%
\pgfsys@useobject{currentmarker}{}%
\end{pgfscope}%
\end{pgfscope}%
\begin{pgfscope}%
\pgfpathrectangle{\pgfqpoint{0.542283in}{0.425000in}}{\pgfqpoint{2.707717in}{1.725000in}}%
\pgfusepath{clip}%
\pgfsetbuttcap%
\pgfsetroundjoin%
\pgfsetlinewidth{0.401500pt}%
\definecolor{currentstroke}{rgb}{0.294118,0.000000,0.509804}%
\pgfsetstrokecolor{currentstroke}%
\pgfsetstrokeopacity{0.850000}%
\pgfsetdash{{1.480000pt}{0.640000pt}}{0.000000pt}%
\pgfpathmoveto{\pgfqpoint{0.665361in}{0.677682in}}%
\pgfusepath{stroke}%
\end{pgfscope}%
\begin{pgfscope}%
\pgfpathrectangle{\pgfqpoint{0.542283in}{0.425000in}}{\pgfqpoint{2.707717in}{1.725000in}}%
\pgfusepath{clip}%
\pgfsetbuttcap%
\pgfsetroundjoin%
\definecolor{currentfill}{rgb}{0.294118,0.000000,0.509804}%
\pgfsetfillcolor{currentfill}%
\pgfsetfillopacity{0.850000}%
\pgfsetlinewidth{1.003750pt}%
\definecolor{currentstroke}{rgb}{0.294118,0.000000,0.509804}%
\pgfsetstrokecolor{currentstroke}%
\pgfsetstrokeopacity{0.850000}%
\pgfsetdash{}{0pt}%
\pgfsys@defobject{currentmarker}{\pgfqpoint{-0.006944in}{-0.006944in}}{\pgfqpoint{0.006944in}{0.006944in}}{%
\pgfpathmoveto{\pgfqpoint{-0.006944in}{-0.006944in}}%
\pgfpathlineto{\pgfqpoint{0.006944in}{0.006944in}}%
\pgfpathmoveto{\pgfqpoint{-0.006944in}{0.006944in}}%
\pgfpathlineto{\pgfqpoint{0.006944in}{-0.006944in}}%
\pgfusepath{stroke,fill}%
}%
\begin{pgfscope}%
\pgfsys@transformshift{0.665361in}{0.677682in}%
\pgfsys@useobject{currentmarker}{}%
\end{pgfscope}%
\end{pgfscope}%
\begin{pgfscope}%
\pgfpathrectangle{\pgfqpoint{0.542283in}{0.425000in}}{\pgfqpoint{2.707717in}{1.725000in}}%
\pgfusepath{clip}%
\pgfsetbuttcap%
\pgfsetroundjoin%
\pgfsetlinewidth{0.401500pt}%
\definecolor{currentstroke}{rgb}{1.000000,0.270588,0.000000}%
\pgfsetstrokecolor{currentstroke}%
\pgfsetstrokeopacity{0.850000}%
\pgfsetdash{{1.480000pt}{0.640000pt}}{0.000000pt}%
\pgfpathmoveto{\pgfqpoint{0.665361in}{2.083348in}}%
\pgfpathlineto{\pgfqpoint{0.802114in}{2.109353in}}%
\pgfpathlineto{\pgfqpoint{1.212374in}{2.110740in}}%
\pgfpathlineto{\pgfqpoint{1.896141in}{2.111730in}}%
\pgfpathlineto{\pgfqpoint{2.579908in}{2.097381in}}%
\pgfpathlineto{\pgfqpoint{3.126922in}{2.097580in}}%
\pgfusepath{stroke}%
\end{pgfscope}%
\begin{pgfscope}%
\pgfpathrectangle{\pgfqpoint{0.542283in}{0.425000in}}{\pgfqpoint{2.707717in}{1.725000in}}%
\pgfusepath{clip}%
\pgfsetbuttcap%
\pgfsetroundjoin%
\definecolor{currentfill}{rgb}{1.000000,0.270588,0.000000}%
\pgfsetfillcolor{currentfill}%
\pgfsetfillopacity{0.850000}%
\pgfsetlinewidth{1.003750pt}%
\definecolor{currentstroke}{rgb}{1.000000,0.270588,0.000000}%
\pgfsetstrokecolor{currentstroke}%
\pgfsetstrokeopacity{0.850000}%
\pgfsetdash{}{0pt}%
\pgfsys@defobject{currentmarker}{\pgfqpoint{-0.006944in}{-0.006944in}}{\pgfqpoint{0.006944in}{0.006944in}}{%
\pgfpathmoveto{\pgfqpoint{-0.006944in}{-0.006944in}}%
\pgfpathlineto{\pgfqpoint{0.006944in}{0.006944in}}%
\pgfpathmoveto{\pgfqpoint{-0.006944in}{0.006944in}}%
\pgfpathlineto{\pgfqpoint{0.006944in}{-0.006944in}}%
\pgfusepath{stroke,fill}%
}%
\begin{pgfscope}%
\pgfsys@transformshift{0.665361in}{2.083348in}%
\pgfsys@useobject{currentmarker}{}%
\end{pgfscope}%
\begin{pgfscope}%
\pgfsys@transformshift{0.802114in}{2.109353in}%
\pgfsys@useobject{currentmarker}{}%
\end{pgfscope}%
\begin{pgfscope}%
\pgfsys@transformshift{1.212374in}{2.110740in}%
\pgfsys@useobject{currentmarker}{}%
\end{pgfscope}%
\begin{pgfscope}%
\pgfsys@transformshift{1.896141in}{2.111730in}%
\pgfsys@useobject{currentmarker}{}%
\end{pgfscope}%
\begin{pgfscope}%
\pgfsys@transformshift{2.579908in}{2.097381in}%
\pgfsys@useobject{currentmarker}{}%
\end{pgfscope}%
\begin{pgfscope}%
\pgfsys@transformshift{3.126922in}{2.097580in}%
\pgfsys@useobject{currentmarker}{}%
\end{pgfscope}%
\end{pgfscope}%
\begin{pgfscope}%
\pgfpathrectangle{\pgfqpoint{0.542283in}{0.425000in}}{\pgfqpoint{2.707717in}{1.725000in}}%
\pgfusepath{clip}%
\pgfsetbuttcap%
\pgfsetroundjoin%
\pgfsetlinewidth{0.401500pt}%
\definecolor{currentstroke}{rgb}{1.000000,0.647059,0.000000}%
\pgfsetstrokecolor{currentstroke}%
\pgfsetstrokeopacity{0.850000}%
\pgfsetdash{{1.480000pt}{0.640000pt}}{0.000000pt}%
\pgfpathmoveto{\pgfqpoint{0.665361in}{0.469231in}}%
\pgfpathlineto{\pgfqpoint{0.802114in}{0.469231in}}%
\pgfpathlineto{\pgfqpoint{1.212374in}{0.469231in}}%
\pgfpathlineto{\pgfqpoint{1.349128in}{0.469231in}}%
\pgfusepath{stroke}%
\end{pgfscope}%
\begin{pgfscope}%
\pgfpathrectangle{\pgfqpoint{0.542283in}{0.425000in}}{\pgfqpoint{2.707717in}{1.725000in}}%
\pgfusepath{clip}%
\pgfsetbuttcap%
\pgfsetroundjoin%
\definecolor{currentfill}{rgb}{1.000000,0.647059,0.000000}%
\pgfsetfillcolor{currentfill}%
\pgfsetfillopacity{0.850000}%
\pgfsetlinewidth{1.003750pt}%
\definecolor{currentstroke}{rgb}{1.000000,0.647059,0.000000}%
\pgfsetstrokecolor{currentstroke}%
\pgfsetstrokeopacity{0.850000}%
\pgfsetdash{}{0pt}%
\pgfsys@defobject{currentmarker}{\pgfqpoint{-0.006944in}{-0.006944in}}{\pgfqpoint{0.006944in}{0.006944in}}{%
\pgfpathmoveto{\pgfqpoint{-0.006944in}{-0.006944in}}%
\pgfpathlineto{\pgfqpoint{0.006944in}{0.006944in}}%
\pgfpathmoveto{\pgfqpoint{-0.006944in}{0.006944in}}%
\pgfpathlineto{\pgfqpoint{0.006944in}{-0.006944in}}%
\pgfusepath{stroke,fill}%
}%
\begin{pgfscope}%
\pgfsys@transformshift{0.665361in}{0.469231in}%
\pgfsys@useobject{currentmarker}{}%
\end{pgfscope}%
\begin{pgfscope}%
\pgfsys@transformshift{0.802114in}{0.469231in}%
\pgfsys@useobject{currentmarker}{}%
\end{pgfscope}%
\begin{pgfscope}%
\pgfsys@transformshift{1.212374in}{0.469231in}%
\pgfsys@useobject{currentmarker}{}%
\end{pgfscope}%
\begin{pgfscope}%
\pgfsys@transformshift{1.349128in}{0.469231in}%
\pgfsys@useobject{currentmarker}{}%
\end{pgfscope}%
\end{pgfscope}%
\begin{pgfscope}%
\pgfpathrectangle{\pgfqpoint{0.542283in}{0.425000in}}{\pgfqpoint{2.707717in}{1.725000in}}%
\pgfusepath{clip}%
\pgfsetbuttcap%
\pgfsetroundjoin%
\pgfsetlinewidth{0.401500pt}%
\definecolor{currentstroke}{rgb}{1.000000,0.498039,0.313725}%
\pgfsetstrokecolor{currentstroke}%
\pgfsetstrokeopacity{0.850000}%
\pgfsetdash{{1.480000pt}{0.640000pt}}{0.000000pt}%
\pgfpathmoveto{\pgfqpoint{0.665361in}{0.469231in}}%
\pgfpathlineto{\pgfqpoint{0.802114in}{0.469231in}}%
\pgfpathlineto{\pgfqpoint{1.212374in}{0.469231in}}%
\pgfpathlineto{\pgfqpoint{1.349128in}{0.469231in}}%
\pgfusepath{stroke}%
\end{pgfscope}%
\begin{pgfscope}%
\pgfpathrectangle{\pgfqpoint{0.542283in}{0.425000in}}{\pgfqpoint{2.707717in}{1.725000in}}%
\pgfusepath{clip}%
\pgfsetbuttcap%
\pgfsetroundjoin%
\definecolor{currentfill}{rgb}{1.000000,0.498039,0.313725}%
\pgfsetfillcolor{currentfill}%
\pgfsetfillopacity{0.850000}%
\pgfsetlinewidth{1.003750pt}%
\definecolor{currentstroke}{rgb}{1.000000,0.498039,0.313725}%
\pgfsetstrokecolor{currentstroke}%
\pgfsetstrokeopacity{0.850000}%
\pgfsetdash{}{0pt}%
\pgfsys@defobject{currentmarker}{\pgfqpoint{-0.006944in}{-0.006944in}}{\pgfqpoint{0.006944in}{0.006944in}}{%
\pgfpathmoveto{\pgfqpoint{-0.006944in}{-0.006944in}}%
\pgfpathlineto{\pgfqpoint{0.006944in}{0.006944in}}%
\pgfpathmoveto{\pgfqpoint{-0.006944in}{0.006944in}}%
\pgfpathlineto{\pgfqpoint{0.006944in}{-0.006944in}}%
\pgfusepath{stroke,fill}%
}%
\begin{pgfscope}%
\pgfsys@transformshift{0.665361in}{0.469231in}%
\pgfsys@useobject{currentmarker}{}%
\end{pgfscope}%
\begin{pgfscope}%
\pgfsys@transformshift{0.802114in}{0.469231in}%
\pgfsys@useobject{currentmarker}{}%
\end{pgfscope}%
\begin{pgfscope}%
\pgfsys@transformshift{1.212374in}{0.469231in}%
\pgfsys@useobject{currentmarker}{}%
\end{pgfscope}%
\begin{pgfscope}%
\pgfsys@transformshift{1.349128in}{0.469231in}%
\pgfsys@useobject{currentmarker}{}%
\end{pgfscope}%
\end{pgfscope}%
\begin{pgfscope}%
\pgfpathrectangle{\pgfqpoint{0.542283in}{0.425000in}}{\pgfqpoint{2.707717in}{1.725000in}}%
\pgfusepath{clip}%
\pgfsetbuttcap%
\pgfsetroundjoin%
\pgfsetlinewidth{0.401500pt}%
\definecolor{currentstroke}{rgb}{0.501961,0.501961,0.000000}%
\pgfsetstrokecolor{currentstroke}%
\pgfsetstrokeopacity{0.850000}%
\pgfsetdash{{1.480000pt}{0.640000pt}}{0.000000pt}%
\pgfpathmoveto{\pgfqpoint{0.665361in}{0.469231in}}%
\pgfpathlineto{\pgfqpoint{0.802114in}{0.513809in}}%
\pgfpathlineto{\pgfqpoint{1.212374in}{0.517092in}}%
\pgfpathlineto{\pgfqpoint{1.378534in}{2.153333in}}%
\pgfusepath{stroke}%
\end{pgfscope}%
\begin{pgfscope}%
\pgfpathrectangle{\pgfqpoint{0.542283in}{0.425000in}}{\pgfqpoint{2.707717in}{1.725000in}}%
\pgfusepath{clip}%
\pgfsetbuttcap%
\pgfsetroundjoin%
\definecolor{currentfill}{rgb}{0.501961,0.501961,0.000000}%
\pgfsetfillcolor{currentfill}%
\pgfsetfillopacity{0.850000}%
\pgfsetlinewidth{1.003750pt}%
\definecolor{currentstroke}{rgb}{0.501961,0.501961,0.000000}%
\pgfsetstrokecolor{currentstroke}%
\pgfsetstrokeopacity{0.850000}%
\pgfsetdash{}{0pt}%
\pgfsys@defobject{currentmarker}{\pgfqpoint{-0.006944in}{-0.006944in}}{\pgfqpoint{0.006944in}{0.006944in}}{%
\pgfpathmoveto{\pgfqpoint{-0.006944in}{-0.006944in}}%
\pgfpathlineto{\pgfqpoint{0.006944in}{0.006944in}}%
\pgfpathmoveto{\pgfqpoint{-0.006944in}{0.006944in}}%
\pgfpathlineto{\pgfqpoint{0.006944in}{-0.006944in}}%
\pgfusepath{stroke,fill}%
}%
\begin{pgfscope}%
\pgfsys@transformshift{0.665361in}{0.469231in}%
\pgfsys@useobject{currentmarker}{}%
\end{pgfscope}%
\begin{pgfscope}%
\pgfsys@transformshift{0.802114in}{0.513809in}%
\pgfsys@useobject{currentmarker}{}%
\end{pgfscope}%
\begin{pgfscope}%
\pgfsys@transformshift{1.212374in}{0.517092in}%
\pgfsys@useobject{currentmarker}{}%
\end{pgfscope}%
\begin{pgfscope}%
\pgfsys@transformshift{1.896141in}{7.250415in}%
\pgfsys@useobject{currentmarker}{}%
\end{pgfscope}%
\begin{pgfscope}%
\pgfsys@transformshift{2.443155in}{7.250415in}%
\pgfsys@useobject{currentmarker}{}%
\end{pgfscope}%
\end{pgfscope}%
\begin{pgfscope}%
\pgfpathrectangle{\pgfqpoint{0.542283in}{0.425000in}}{\pgfqpoint{2.707717in}{1.725000in}}%
\pgfusepath{clip}%
\pgfsetbuttcap%
\pgfsetroundjoin%
\pgfsetlinewidth{0.401500pt}%
\definecolor{currentstroke}{rgb}{0.235294,0.701961,0.443137}%
\pgfsetstrokecolor{currentstroke}%
\pgfsetstrokeopacity{0.850000}%
\pgfsetdash{{1.480000pt}{0.640000pt}}{0.000000pt}%
\pgfpathmoveto{\pgfqpoint{0.665361in}{0.675561in}}%
\pgfpathlineto{\pgfqpoint{0.802114in}{0.570146in}}%
\pgfpathlineto{\pgfqpoint{1.212374in}{0.571564in}}%
\pgfpathlineto{\pgfqpoint{1.896141in}{0.570137in}}%
\pgfpathlineto{\pgfqpoint{2.579908in}{0.573688in}}%
\pgfpathlineto{\pgfqpoint{2.853415in}{0.561608in}}%
\pgfusepath{stroke}%
\end{pgfscope}%
\begin{pgfscope}%
\pgfpathrectangle{\pgfqpoint{0.542283in}{0.425000in}}{\pgfqpoint{2.707717in}{1.725000in}}%
\pgfusepath{clip}%
\pgfsetbuttcap%
\pgfsetroundjoin%
\definecolor{currentfill}{rgb}{0.235294,0.701961,0.443137}%
\pgfsetfillcolor{currentfill}%
\pgfsetfillopacity{0.850000}%
\pgfsetlinewidth{1.003750pt}%
\definecolor{currentstroke}{rgb}{0.235294,0.701961,0.443137}%
\pgfsetstrokecolor{currentstroke}%
\pgfsetstrokeopacity{0.850000}%
\pgfsetdash{}{0pt}%
\pgfsys@defobject{currentmarker}{\pgfqpoint{-0.006944in}{-0.006944in}}{\pgfqpoint{0.006944in}{0.006944in}}{%
\pgfpathmoveto{\pgfqpoint{-0.006944in}{-0.006944in}}%
\pgfpathlineto{\pgfqpoint{0.006944in}{0.006944in}}%
\pgfpathmoveto{\pgfqpoint{-0.006944in}{0.006944in}}%
\pgfpathlineto{\pgfqpoint{0.006944in}{-0.006944in}}%
\pgfusepath{stroke,fill}%
}%
\begin{pgfscope}%
\pgfsys@transformshift{0.665361in}{0.675561in}%
\pgfsys@useobject{currentmarker}{}%
\end{pgfscope}%
\begin{pgfscope}%
\pgfsys@transformshift{0.802114in}{0.570146in}%
\pgfsys@useobject{currentmarker}{}%
\end{pgfscope}%
\begin{pgfscope}%
\pgfsys@transformshift{1.212374in}{0.571564in}%
\pgfsys@useobject{currentmarker}{}%
\end{pgfscope}%
\begin{pgfscope}%
\pgfsys@transformshift{1.896141in}{0.570137in}%
\pgfsys@useobject{currentmarker}{}%
\end{pgfscope}%
\begin{pgfscope}%
\pgfsys@transformshift{2.579908in}{0.573688in}%
\pgfsys@useobject{currentmarker}{}%
\end{pgfscope}%
\begin{pgfscope}%
\pgfsys@transformshift{2.853415in}{0.561608in}%
\pgfsys@useobject{currentmarker}{}%
\end{pgfscope}%
\end{pgfscope}%
\begin{pgfscope}%
\pgfpathrectangle{\pgfqpoint{0.542283in}{0.425000in}}{\pgfqpoint{2.707717in}{1.725000in}}%
\pgfusepath{clip}%
\pgfsetbuttcap%
\pgfsetroundjoin%
\pgfsetlinewidth{0.401500pt}%
\definecolor{currentstroke}{rgb}{0.501961,0.501961,0.501961}%
\pgfsetstrokecolor{currentstroke}%
\pgfsetstrokeopacity{0.850000}%
\pgfsetdash{{1.480000pt}{0.640000pt}}{0.000000pt}%
\pgfpathmoveto{\pgfqpoint{0.665361in}{0.469231in}}%
\pgfpathlineto{\pgfqpoint{0.802114in}{0.469231in}}%
\pgfpathlineto{\pgfqpoint{1.212374in}{0.469231in}}%
\pgfpathlineto{\pgfqpoint{1.485881in}{0.469231in}}%
\pgfusepath{stroke}%
\end{pgfscope}%
\begin{pgfscope}%
\pgfpathrectangle{\pgfqpoint{0.542283in}{0.425000in}}{\pgfqpoint{2.707717in}{1.725000in}}%
\pgfusepath{clip}%
\pgfsetbuttcap%
\pgfsetroundjoin%
\definecolor{currentfill}{rgb}{0.501961,0.501961,0.501961}%
\pgfsetfillcolor{currentfill}%
\pgfsetfillopacity{0.850000}%
\pgfsetlinewidth{1.003750pt}%
\definecolor{currentstroke}{rgb}{0.501961,0.501961,0.501961}%
\pgfsetstrokecolor{currentstroke}%
\pgfsetstrokeopacity{0.850000}%
\pgfsetdash{}{0pt}%
\pgfsys@defobject{currentmarker}{\pgfqpoint{-0.006944in}{-0.006944in}}{\pgfqpoint{0.006944in}{0.006944in}}{%
\pgfpathmoveto{\pgfqpoint{-0.006944in}{-0.006944in}}%
\pgfpathlineto{\pgfqpoint{0.006944in}{0.006944in}}%
\pgfpathmoveto{\pgfqpoint{-0.006944in}{0.006944in}}%
\pgfpathlineto{\pgfqpoint{0.006944in}{-0.006944in}}%
\pgfusepath{stroke,fill}%
}%
\begin{pgfscope}%
\pgfsys@transformshift{0.665361in}{0.469231in}%
\pgfsys@useobject{currentmarker}{}%
\end{pgfscope}%
\begin{pgfscope}%
\pgfsys@transformshift{0.802114in}{0.469231in}%
\pgfsys@useobject{currentmarker}{}%
\end{pgfscope}%
\begin{pgfscope}%
\pgfsys@transformshift{1.212374in}{0.469231in}%
\pgfsys@useobject{currentmarker}{}%
\end{pgfscope}%
\begin{pgfscope}%
\pgfsys@transformshift{1.485881in}{0.469231in}%
\pgfsys@useobject{currentmarker}{}%
\end{pgfscope}%
\end{pgfscope}%
\begin{pgfscope}%
\pgfpathrectangle{\pgfqpoint{0.542283in}{0.425000in}}{\pgfqpoint{2.707717in}{1.725000in}}%
\pgfusepath{clip}%
\pgfsetbuttcap%
\pgfsetroundjoin%
\pgfsetlinewidth{0.401500pt}%
\definecolor{currentstroke}{rgb}{0.000000,0.501961,0.501961}%
\pgfsetstrokecolor{currentstroke}%
\pgfsetstrokeopacity{0.850000}%
\pgfsetdash{{1.480000pt}{0.640000pt}}{0.000000pt}%
\pgfpathmoveto{\pgfqpoint{0.665361in}{0.469231in}}%
\pgfpathlineto{\pgfqpoint{0.802114in}{0.469231in}}%
\pgfpathlineto{\pgfqpoint{1.212374in}{0.469231in}}%
\pgfpathlineto{\pgfqpoint{1.896141in}{0.469231in}}%
\pgfpathlineto{\pgfqpoint{2.032895in}{0.469231in}}%
\pgfusepath{stroke}%
\end{pgfscope}%
\begin{pgfscope}%
\pgfpathrectangle{\pgfqpoint{0.542283in}{0.425000in}}{\pgfqpoint{2.707717in}{1.725000in}}%
\pgfusepath{clip}%
\pgfsetbuttcap%
\pgfsetroundjoin%
\definecolor{currentfill}{rgb}{0.000000,0.501961,0.501961}%
\pgfsetfillcolor{currentfill}%
\pgfsetfillopacity{0.850000}%
\pgfsetlinewidth{1.003750pt}%
\definecolor{currentstroke}{rgb}{0.000000,0.501961,0.501961}%
\pgfsetstrokecolor{currentstroke}%
\pgfsetstrokeopacity{0.850000}%
\pgfsetdash{}{0pt}%
\pgfsys@defobject{currentmarker}{\pgfqpoint{-0.006944in}{-0.006944in}}{\pgfqpoint{0.006944in}{0.006944in}}{%
\pgfpathmoveto{\pgfqpoint{-0.006944in}{-0.006944in}}%
\pgfpathlineto{\pgfqpoint{0.006944in}{0.006944in}}%
\pgfpathmoveto{\pgfqpoint{-0.006944in}{0.006944in}}%
\pgfpathlineto{\pgfqpoint{0.006944in}{-0.006944in}}%
\pgfusepath{stroke,fill}%
}%
\begin{pgfscope}%
\pgfsys@transformshift{0.665361in}{0.469231in}%
\pgfsys@useobject{currentmarker}{}%
\end{pgfscope}%
\begin{pgfscope}%
\pgfsys@transformshift{0.802114in}{0.469231in}%
\pgfsys@useobject{currentmarker}{}%
\end{pgfscope}%
\begin{pgfscope}%
\pgfsys@transformshift{1.212374in}{0.469231in}%
\pgfsys@useobject{currentmarker}{}%
\end{pgfscope}%
\begin{pgfscope}%
\pgfsys@transformshift{1.896141in}{0.469231in}%
\pgfsys@useobject{currentmarker}{}%
\end{pgfscope}%
\begin{pgfscope}%
\pgfsys@transformshift{2.032895in}{0.469231in}%
\pgfsys@useobject{currentmarker}{}%
\end{pgfscope}%
\end{pgfscope}%
\begin{pgfscope}%
\pgfsetrectcap%
\pgfsetmiterjoin%
\pgfsetlinewidth{0.803000pt}%
\definecolor{currentstroke}{rgb}{0.000000,0.000000,0.000000}%
\pgfsetstrokecolor{currentstroke}%
\pgfsetdash{}{0pt}%
\pgfpathmoveto{\pgfqpoint{0.542283in}{0.425000in}}%
\pgfpathlineto{\pgfqpoint{0.542283in}{2.150000in}}%
\pgfusepath{stroke}%
\end{pgfscope}%
\begin{pgfscope}%
\pgfsetrectcap%
\pgfsetmiterjoin%
\pgfsetlinewidth{0.803000pt}%
\definecolor{currentstroke}{rgb}{0.000000,0.000000,0.000000}%
\pgfsetstrokecolor{currentstroke}%
\pgfsetdash{}{0pt}%
\pgfpathmoveto{\pgfqpoint{3.250000in}{0.425000in}}%
\pgfpathlineto{\pgfqpoint{3.250000in}{2.150000in}}%
\pgfusepath{stroke}%
\end{pgfscope}%
\begin{pgfscope}%
\pgfsetrectcap%
\pgfsetmiterjoin%
\pgfsetlinewidth{0.803000pt}%
\definecolor{currentstroke}{rgb}{0.000000,0.000000,0.000000}%
\pgfsetstrokecolor{currentstroke}%
\pgfsetdash{}{0pt}%
\pgfpathmoveto{\pgfqpoint{0.542283in}{0.425000in}}%
\pgfpathlineto{\pgfqpoint{3.250000in}{0.425000in}}%
\pgfusepath{stroke}%
\end{pgfscope}%
\begin{pgfscope}%
\pgfsetrectcap%
\pgfsetmiterjoin%
\pgfsetlinewidth{0.803000pt}%
\definecolor{currentstroke}{rgb}{0.000000,0.000000,0.000000}%
\pgfsetstrokecolor{currentstroke}%
\pgfsetdash{}{0pt}%
\pgfpathmoveto{\pgfqpoint{0.542283in}{2.150000in}}%
\pgfpathlineto{\pgfqpoint{3.250000in}{2.150000in}}%
\pgfusepath{stroke}%
\end{pgfscope}%
\end{pgfpicture}%
\makeatother%
\endgroup%

%% file: content/09-related.tex
\section{Related Work \label{sec:related}}
Several incremental algorithms for particular data-flow problems have been proposed since the 1980s.
Ryder \cite{DBLP:conf/popl/Ryder83} presents incremental algorithms for certain forward and backward data flow algorithms.
Zadeck \cite{DBLP:conf/sigplan/Zadeck84} proposes an incremental data flow analysis, relying on specific handling depending on the type of change,
such as adding variable definitions or adding or removing edges.
Further work from that time on how to make data-flow analyses incremental includes \cite{DBLP:conf/popl/CarrollR88,DBLP:conf/icse/YurRLS97,DBLP:conf/sigplan/CooperK84,DBLP:journals/sigplan/CooperK88}, while
Arzt and Bodden \cite{DBLP:conf/icse/ArztB14} report on more recent results.

Other work focuses on how to make static program analysis techniques other than data-flow analysis more responsive and/or incremental.
Do et al.~\cite{DBLP:conf/issta/DoALBSM17} describe a layered approach, where static analysis is first performed locally,
e.g.\ intra-procedurally, to produce some analysis results fast, and then proceeds to analyze bigger layers of code.
While this allows to quickly present parts of the feedback to the developer, this approach does not speed up whole-program analysis.

Modular, summary-based static analyses \cite{cousot2001compositional} naturally lend themselves for incrementalization.
Van der Plas et al.\ \cite{DBLP:conf/scam/PlasSER20} propose a modular flow analysis framework that they instantiate with
function-modular and thread-modular analyses of \textsc{Scheme} code.
Their analysis builds on a worklist algorithm.
For an incremental analysis run, only those components of the analysis are initially added to the working list which are \textit{directly} influenced by the program change.
Indirectly affected components are added only during the execution of the worklist algorithm.
The proposed algorithm cannot improve on the previous results, as old and new abstract values are always joined.
The addition of new variable bindings is handled by introducing a
dummy declaration of the variable into the \textit{old} program version.
This limits the practical applicability where the changes to be performed are not known beforehand.

\textsc{Infer} \cite{DBLP:conf/nfm/CalcagnoD11,infer,DBLP:conf/lics/OHearn18}, a compositional static analysis tool, is used in the development
process inside \textsc{Facebook}.
\textsc{Infer} supports an incremental analysis mode where only changed functions and functions affected by changed functions are reanalyzed.
The speed-up obtained by the incremental analysis is utilized to obtain faster feedback in the development process,
as incremental runs are performed when a developer submits a pull-request for peer-review.
An issue shared with any summary-based approach is the expressivity of the formalism for expressing summaries.
So the lack of context-sensitivity can deteriorate analysis precision.

McPeak et al.~\cite{DBLP:conf/sigsoft/McPeakGR13} describe an approach for incremental and parallel bug-detection
used in the commercial tool \textsc{Coverity}.
The framework has been evaluated on large code bases, but is not claimed to be sound.

As mentioned in Section~\ref{sec:introduction},
there is some recent work on making analyses based on abstract interpretation incremental.
In \cite{Stein21}, abstract interpretation is performed using an explicitly built acyclic \emph{demanded abstract interpretation graph}.
In this graph, the representation of loops has to be unrolled until a fixpoint is reached.
This means that intermediate results for unknowns relating to program points in loops are maintained, which is not necessary in our framework.
The incremental analysis there guarantees from-scratch-consistency.
An extension of the work, describing how recursive functions and dynamic function calls can be handled, is provided in \cite{stein_phd}.
Opposed to our work, the experimental evaluation in \cite{Stein21} is performed on a synthetic benchmark where each program change consists of the addition
of a single statement, loop or if-then-else conditional.
It is not clear how a complete reanalysis can be avoided when, e.g., initializers of global program variables are changed.
Further, it is not shown how this framework could handle downward iterations with narrowing or deal with multithreaded code.

A version of the top-down solver is used in \cite{DBLP:journals/toplas/HermenegildoPMS00,DBLP:journals/tplp/Garcia-Contreras21} for an incremental analysis of code using constrained Horn clauses.
In this setting, the incrementality is \linebreak[4] achieved by a specific handling for addition and removal of Horn clauses.
Another version of the top-down solver (now with side-effects as ours) is considered in \cite{Seidl2020}. That paper concentrates on the
minimal requirements for turning this demand-driven solver into an incremental one.
No effort is made to speed up efficiency of reanalysis by dedicated means such as reluctant destabilization or incremental postprocessing.
Neither do they consider means for retaining precision for context-insensitive unknowns, e.g., by restarting.
In fact, restarting of certain unknowns has been proposed by \cite{Halbwachs12} to
improve the precision of the analysis of a single program version in the presence of widening.
In contrast, restarting in our context is targeted at alleviating the precision-loss
caused by accumulated abstract values of flow-insensitive unknowns across multiple reanalysis runs.

%% file: content/10-conclusion.tex
\section{Conclusion}
\label{sec:conclusion}

We have presented a framework for interactive static analysis which puts abstract interpretation at the fingertips of the
developer. For abstract interpretation, we relied on the analyzer \textsc{Goblint}, while \textsc{MagpieBridge} serves as the interface to
IDEs.
We indicated that, in order to obtain acceptable response times, the whole analysis pipeline should be incrementalized.
We presented techniques to achieve this, namely,
incrementalization of the underlying solver, reluctant destabilization, and restarting.
We demonstrated that also incrementalizing the postprocessing phase greatly helped to reduce reanalysis times.
For the given benchmark repository, these methods sufficed.
We expect, however, that for larger project repositories, \emph{preprocessing} also need to be incremental.

Besides \emph{efficiency} as a yardstick for incremental reanalysis, we considered \emph{consistency} of the computed results
as well as \emph{usability} of the overall approach. While achieving a speedup by a factor of roughly 40 in the majority of cases (compared to
from-scratch analysis), we found the reanalysis results to be surprisingly often consistent.
Concerning usability, we provided meaningful stories which indicate that our setup indeed can successfully be used in a close \emph{analyze-modify}
loop.
Currently, just warnings are reported back from the analyzer to the IDE.
This may suffice in many cases, since it does not require any extra background knowledge from the developer.
For future work, we would like to explore which further hints or insights into the program we should provide to developers to
understand analysis results and support them in realizing high quality code.
The analysis result itself may be too complex to be of any help.
Further, one may explore how the approaches presented here can be used to obtain an analysis that remains
scalable and thus usable for large scale projects, while allowing to analyze critical parts of the code with higher precision, in order
to lower the burden of code inspection and thus increase user acceptance.
%